\documentclass[a4paper,11pt]{article}
\usepackage[utf8]{inputenc}
\usepackage[T1]{fontenc}
\usepackage{amsmath}
\usepackage{epsfig}
\usepackage{epic}
\usepackage{setspace}

\newtheorem{theorem}{Theorem}

\newtheorem{proposition}[theorem]{Proposition}
\newenvironment{proof}[1][Proof]{\noindent \textbf{#1.} }{\  \rule{0.5em}{0.5em}}
\newcommand{\dlim}{\xrightarrow{\cal L}}

\begin{document}

\title{Confidence distributions for the autoregressive parameter}
\author{Rolf Larsson\footnote{Dept of Mathematics, Uppsala University, P.O.Box 480, SE-751 06 Uppsala, Sweden, rolf.larsson@math.uu.se.}}
\maketitle

\begin{abstract}
The notion of confidence distributions is applied to inference about the parameter in a simple autoregressive model, allowing the parameter to take the value one. This makes it possible to compare to asymptotic approximations in both the stationary and the non stationary cases at the same time. The main point, however, is to compare to a Bayesian analysis of the same problem. A non informative prior for a parameter, in the sense of Jeffreys, is given as the ratio of the confidence density and the likelihood. In this way, the similarity between the confidence and non-informative Bayesian frameworks is exploited. It is shown that, in the stationary case, asymptotically the so induced prior is flat. However, if a unit parameter is allowed, the induced prior has to have a spike at one of some size.
Simulation studies and two empirical examples illustrate the ideas. 
\end{abstract}

\newpage

\section{Introduction}

Historically, Laplace and others conducted inference on population parameters based on the notion of inverse probability, grounded in Bayes' formula and the assumption of a flat prior, see e.g. Hald (2007). Fisher (1930) challenged this view, and introduced fiducial probability, making it possible to formulate probability statements on a parameter without imposing any prior distribution. Later on, this approach has been heavily criticized, see e.g. Neyman (1941), who instead put forward the concept of confidence. He argued that, other than in some particular cases, confidence is not equivalent to the fiducial approach. Cox (1958) discussed these ideas and introduced the term confidence distributions.

Similar to the fiducial setup, the methodology of confidence distributions constructs probability distributions in an epistemic sense for parameters  without incorporating any prior beliefs. In this way, many attractive features can be gained, without having to impose too many subjective elements into the statistical analysis. In recent years, these methods have enjoyed new popularity, see e.g. the excellent monograph Schweder and Hjort (2016), or the review paper Xie and Singh (2013).

Our paper deals with a simple autoregressive model of order one. Fairly recent studies of constructing confidence intervals in autoregressive models, obtained by inverting hypothesis tests and using local to unit asymptotics, are Elliott and Stock (2001) and Phillips (2014). 

In the present paper, we consider the confidence distribution for the autoregressive parameter. The main question to answer is how the corresponding confidence density relates to a Bayesian posterior. Which prior distribution makes them equal? In particular, for nearly non stationary cases, is there a spike at one in this prior distribution?

The rest of the paper is as follows. Section 2 lays out some relevant facts on confidence distributions. In section 3, we apply the concepts to the order one autoregressive process, illustrated with simulations. Section 4 presents two empirical examples  and section 5 concludes.

All numerical calculations of the paper have been performed using Matlab 2019b and R. Program codes may be obtained at request.

\section{Confidence distributions}

Following Schweder and Hjort (2016), the idea of confidence distributions is as follows. Assume that we have a statistical model with one unknown scalar parameter $\theta$. We have observations $y_1,...,y_n$ and a maximum likelihood estimate $\hat\theta_{obs}$ that is a function of these observations. For a given $\theta=\theta_0$, we want to test $H_0$: $\theta\leq\theta_0$ vs $H_1$: $\theta>\theta_0$.
If we base the test on $\hat\theta_{obs}$ and reject $H_0$ for large values, then the p value of the test is given by
\begin{equation}
C(\theta_0)=P_{\theta_0}(\hat\theta\geq\hat\theta_{obs}),\label{cdf}
\end{equation}
where we calculate the probability under $\theta=\theta_0$, and where $\hat\theta$ is the random variable of which $\hat\theta_{obs}$ is an observation.

The p value in (\ref{cdf}) expresses our confidence in $H_0$. Given $\hat\theta_{obs}$, if $\theta_0$ is small, the random variable $\hat\theta$ tends to be small, and the p value $C(\theta_0)$ is close to zero. Conversely, for large $\theta_0$, the p value is close to one. It is reasonable that, just as a cumulative distribution function, $C(\theta_0)$ is a monotone increasing function of $\theta_0$. In this section, from here and on we only consider continuous distributions.
We say that $C(\theta_0)$ is the {\it confidence distribution function} of the parameter $\theta_0$.

In fact, if we let $\alpha\in(0,1)$ be given, and define $\theta=\theta_1$ so that\\ $C(\theta_0)\leq 1-\alpha$ for all $\theta_0\leq\theta_1$ and $C(\theta_1)=1-\alpha$, then a one sided confidence interval for $\theta$ with confidence degree $1-\alpha$ is given by $(-\infty,\theta_1)$. From this, it follows that under the distribution of $\hat\theta$, if $C(\theta)$ is strictly increasing in $\theta$, then
$$1-\alpha=P(\theta\leq\theta_1)=P\{C(\theta)\leq C(\theta_1)\}
=P\{C(\theta)\leq 1-\alpha\}.$$
Hence, $C(\theta)$ follows the uniform distribution on $(0,1)$. This can be taken as the formal definition of a confidence distribution, see Schweder and Hjort (2016), p.58 and Xie and Singh (2013).

Here, we also want to mention a related concept, the {\it confidence curve} (Birnbaum, 1961), which is defined as
\begin{equation}
cc(\theta)=|1-2C(\theta)|.\label{cc}
\end{equation}
It is easily seen that all $\theta$ satisfying $cc(\theta)\leq 1-\alpha$ form an equally tailed $1-\alpha$ confidence interval for $\theta$. Hence, the confidence curve depicts equally tailed confidence intervals for all confidence levels at the same time.

Viewing $C(\theta)$ as a confidence distribution function, it is obvious that we can also define the corresponding confidence density function by differentiating $C(\theta)$ with respect to $\theta$, i.e we define the confidence density function of $\theta$ as
\begin{equation}
c(\theta)=\frac{d}{d\theta}C(\theta).\label{df}
\end{equation}
The confidence density function in (\ref{df}) is interesting, because it can be compared to the posterior density of $\theta$ in a Bayesian setup. In fact, if $f(y|\theta)$ is the likelihood function given $\theta$, then we can find an implied prior density $h(\theta)$ for $\theta$ through
\begin{equation}
c(\theta)=f(y|\theta)h(\theta),\label{c}
\end{equation}
where the confidence density plays the role of a Bayesian posterior.

Lindley (1958) has shown that in univariate location parameter problems, the prior satisfying (\ref{c}) is the Jeffrey's prior (Jeffreys, 1931). In the following, we will refer to such a prior as being non informative for $\theta$. This is in the sense that it preserves the posterior distribution in the presence of parameter transformations. It does not mean that the prior has to be flat, cf. the Jeffreys prior for the normal standard deviation $\sigma$, being $1/\sigma$, see Jeffreys (1946). 

\section{The autoregressive model}

\subsection{Confidence distribution function}

Consider an autoregressive process of order one,
\begin{equation}
Y_t=\phi Y_{t-1}+\varepsilon_t,\label{AR1}
\end{equation}
where $t=1,2,...$ and the $\varepsilon_t$ are independent and normally distributed with expectation 0 and unknown variance $\sigma^2$, $N(0,\sigma^2)$. Say that we have observations $y_1,...,y_n$.  

We may find the confidence distribution for $\phi$ by following the general recipe of the previous section. Assuming $Y_0=0$, then by analogy to simple no intercept linear regression, the maximum likelihood estimator of $\phi$ is given by
\begin{equation}
\hat\phi=\frac{\sum_{t=1}^n Y_{t-1}Y_t}{\sum_{t=1}^n Y_{t-1}^2}.
\label{MLE}
\end{equation}
Now, the sample gives rise to the corresponding estimate $\hat\phi_{obs}$, and testing $H_0$: $\phi\leq\phi_0$ vs $H_1$: $\phi>\phi_0$ gives rise to the confidence distribution
\begin{equation}
C(\phi)=P_{\phi}(\hat\phi\geq\hat\phi_{obs}).\label{arcdf}
\end{equation}
For finite samples, we are not able to find an explicit expression for (\ref{arcdf}), but it may be readily simulated. The simulation setup is described in Appendix 1.

For a suitable choice of initial distribution, the process in (\ref{AR1}) is stationary if and only if $|\phi|<1$. For such a $\phi$, the asymptotic distribution of $\sqrt{n}(\hat\phi-\phi)$ is $N(0,1-\phi^2)$, sse e.g. Shumway and Stoffer (2017), p.126.
This leads to the statistic
$\sqrt{n}(\hat\phi-\phi)/\sqrt{1-\phi^2}$, which is asymptotically standard normal. Thus, for large $n$,  we get, denoting the standard normal cumulative distribution function by $\Phi(\cdot)$,
\begin{align}
C(\phi)&=P_{\phi}(\hat\phi\geq\hat\phi_{obs})=1-P_{\phi}(\hat\phi<\hat\phi_{obs})\notag\\
&\approx 1-\Phi\left(\sqrt{n}\frac{\hat\phi_{obs}-\phi}{\sqrt{1-\phi^2}}\right)
=\Phi\left(\sqrt{n}\frac{\phi-\hat\phi_{obs}}{\sqrt{1-\phi^2}}\right)
=C_1(\phi),\label{asarcdf1}
\end{align}
where $C_1(\phi)$ is an approximate confidence distribution for large $n$ and $|\phi|<1$.

In the variance expression, it is also possible to insert $\hat\phi_{obs}$ for $\phi$ , so then we start with the asymptotic standard normal statistic $\sqrt{n}(\hat\phi-\phi)/\sqrt{1-\hat\phi_{obs}^2}$. As in (\ref{asarcdf1}), this leads to a new approximate confidence distribution function as
\begin{equation}
C_2(\phi)=\Phi\left(\sqrt{n}\frac{\phi-\hat\phi_{obs}}{\sqrt{1-\hat\phi_{obs}^2}}\right).\label{asarcdf2}
\end{equation}

Considering $\phi=1$, stationarity is violated and we have the asymptotic result
\begin{equation}
n(\hat\phi-1)\dlim\frac{\int_0^1 W(t)dW(t)}{\int_0^1W(t)^2dt}=\frac{W(1)^2-1}{2\int_0^1W(t)^2dt}=Z,\label{Z}
\end{equation}
as $n\to\infty$, where $\dlim$ denotes convergence in law (distribution), and where $\{W(t)\}$ is a standard Wiener process. The distribution of the quantity $Z$ on the right hand side of (\ref{Z}), called the Dickey-Fuller distribution, is non standard and needs to be simulated. In fact, (\ref{Z}) is a special case of the near unit root asymptotic result (Phillips, 1987, 2014, Perron, 1989)
\begin{equation}
n(\hat\phi-\phi)\dlim\frac{\int_0^1 J_c(t)dW(t)}{\int_0^1J_c(t)^2dt},\label{OU}
\end{equation}
where for a constant $c$, 
\begin{equation}
\phi=\exp(c/n)=1+c/n+O(n^{-2})\label{phic}
\end{equation}
and
$$J_c(t)=\int_0^t e^{(t-s)c}dW(s)$$
is an Ornstein-Uhlenbeck process. The stationary case corresponds to $c<0$.

If we define $F(\cdot)$ as the cumulative distribution function of $Z$, it follows as in (\ref{asarcdf1}) that, evaluating probabilities at $\phi=1$, for large $n$,
\begin{equation}
C(1)=P(\hat\phi\geq\hat\phi_{obs})=1-P(\hat\phi<\hat\phi_{obs})
\approx 1-F\left\{n(\hat\phi_{obs}-1)\right\},\label{asurcdf}
\end{equation}
where the right hand side gives an approximate confidence distribution for large $n$ and $\phi=1$.
Note that our confidence distribution function in (\ref{asurcdf}) corresponds to the hypotheses $H_0$: $\phi=1$, $H_1$: $\phi>1$, while the unit root test usually has $H_1$: $\phi<1$ as alternative. This corresponds to the fact that the asymptotic confidence density function in (\ref{asurcdf}) is of the form $1-F$, not $F$ as for the usual test formulation.

Figures 1-2 below show confidence distributions obtained by simulation, cf (\ref{arcdf}) and Appendix 1, by asymptotic approximations from (\ref{asarcdf1}) and (\ref{asarcdf2}). Inspired by (\ref{OU}) and (\ref{phic}), we put $\hat\phi_{obs}=1-c/n$ , with $c=5, 10$ and we have $n=100, 200, 400$. The number of replications is $10^4$. For $\phi=1$, we simulated from the asymptotic approximation (\ref{Z}) with $10^6$ replicates. In all cases, the result was basically the same as the one obtained from (\ref{arcdf}).
We find that the shapes of the graphs are very similar for the same $c$. The values at 1 are also similar, at about 0.875 for $c=5$ and 0.972 for $c=10$, which agrees very well with the corresponding values obtained by numerical methods in Perron (1989). 

Moreover, note that the approximations are closer to the simulations for smaller $\hat\phi_{obs}$. This is expected, since the asymptotic normal approximation breaks down for $\phi$ values (that generated $\hat\phi_{obs}$) close to one. Another important observation is that overall, the approximation in (\ref{asarcdf2}) works better than the one in (\ref{asarcdf1}). Intuitively, this should be because in (\ref{asarcdf1}), the asymptotic variance of $\sqrt{n}\hat\phi$ is computed as $1-\phi^2$, but note that $\phi$ is just the argument of $C(\phi)$. It does not in general equal the true parameter, $\phi_0$ say, that is thought to have generated $\hat\phi_{obs}$. Supposedly, the variance approximation of (\ref{asarcdf2}), $1-\hat\phi_{obs}^2$, should come closer to the true variance $1-\phi_0^2$.

Confidence intervals may be read off from confidence distributions. For example, an equally tailed $1-\alpha$ confidence interval for the parameter is given by $(\phi_1,\phi_2)$ where $C(\phi_1)=\alpha/2$ and $C(\phi_2)=1-(\alpha/2)$.  
This is with the modification that $C(\phi)$ may be improper in the sense that it never reaches the value $1-(\alpha/2)$ in the parameter range of $\phi$. In this case the rightmost part of the confidence curve never reaches $1-\alpha$. Instead, the upper bound of the interval equals one, which is the upper bound of the parameter space.

The latter is more easily seen from the empirical confidence curve of Figure 3. This curve is based on the same simulations as above, for $\hat\phi_{obs}=0.90$ and $n=100$, where the 90\% and 95\% levels are emphasized. The corresponding confidence intervals for $\phi$ can be read off at the $\phi$ axis (or for better precision, numerically solved) to be about $(0.833,\ 0.986)$ and $(0.819,\ 1)$, respectively. Also, observe that the confidence curve `points at' about $0.909$. This is the median of the confidence distribution, and it can be considered as an alternative estimate of $\phi$.

\begin{figure}[htb]
\begin{center}
\includegraphics[width=125mm,height=50mm]{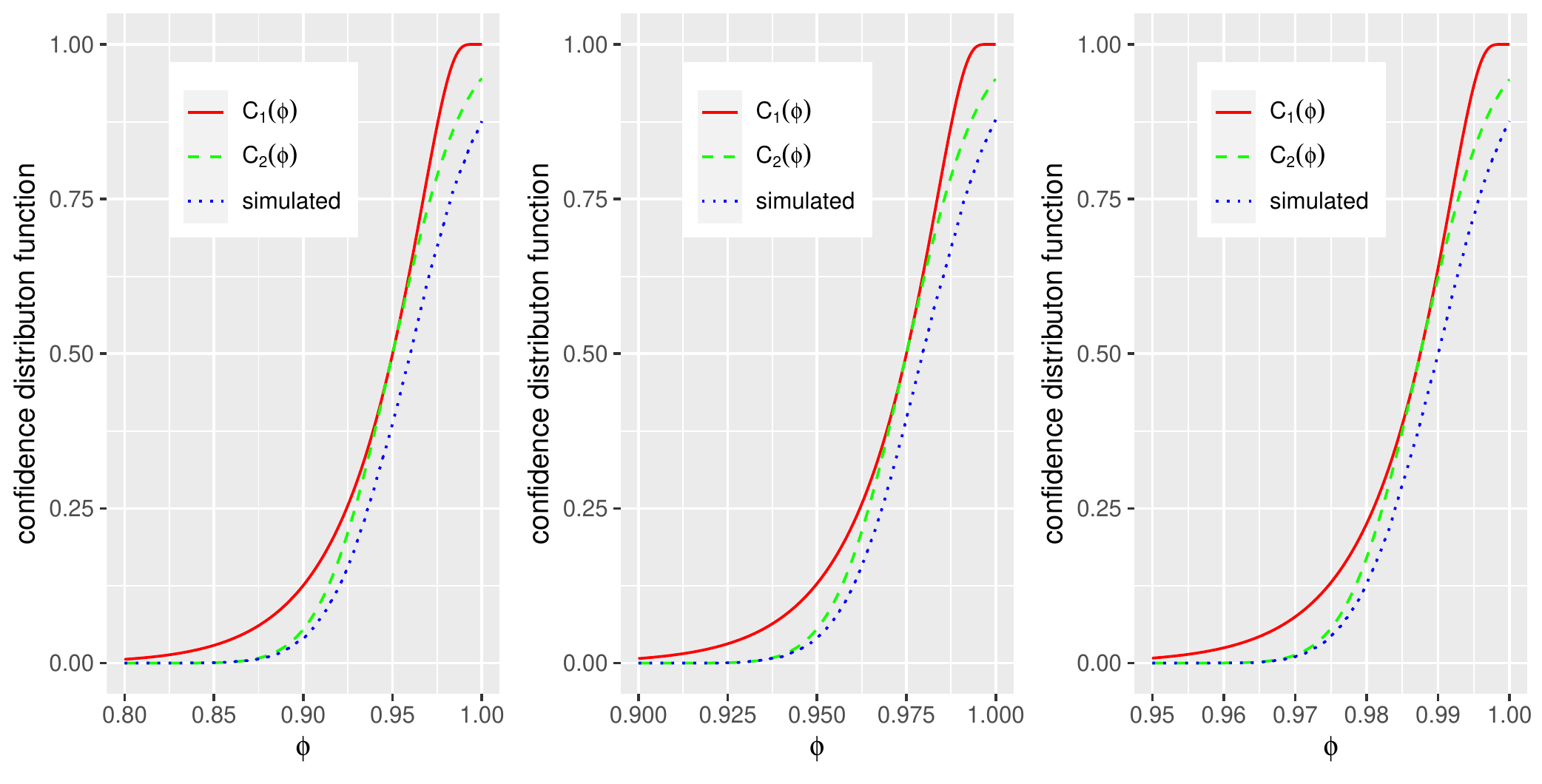}
\end{center}
\caption{Simulated and asymptotic confidence distributions, $n=100, 200, 400$ and $\hat\phi_{obs}=0.95, 0.975, 0.9875$ (i.e. $c=5$ in (\ref{phic})) from left to right.}
\end{figure}

\begin{figure}[htb]
\begin{center}
\includegraphics[width=125mm,height=50mm]{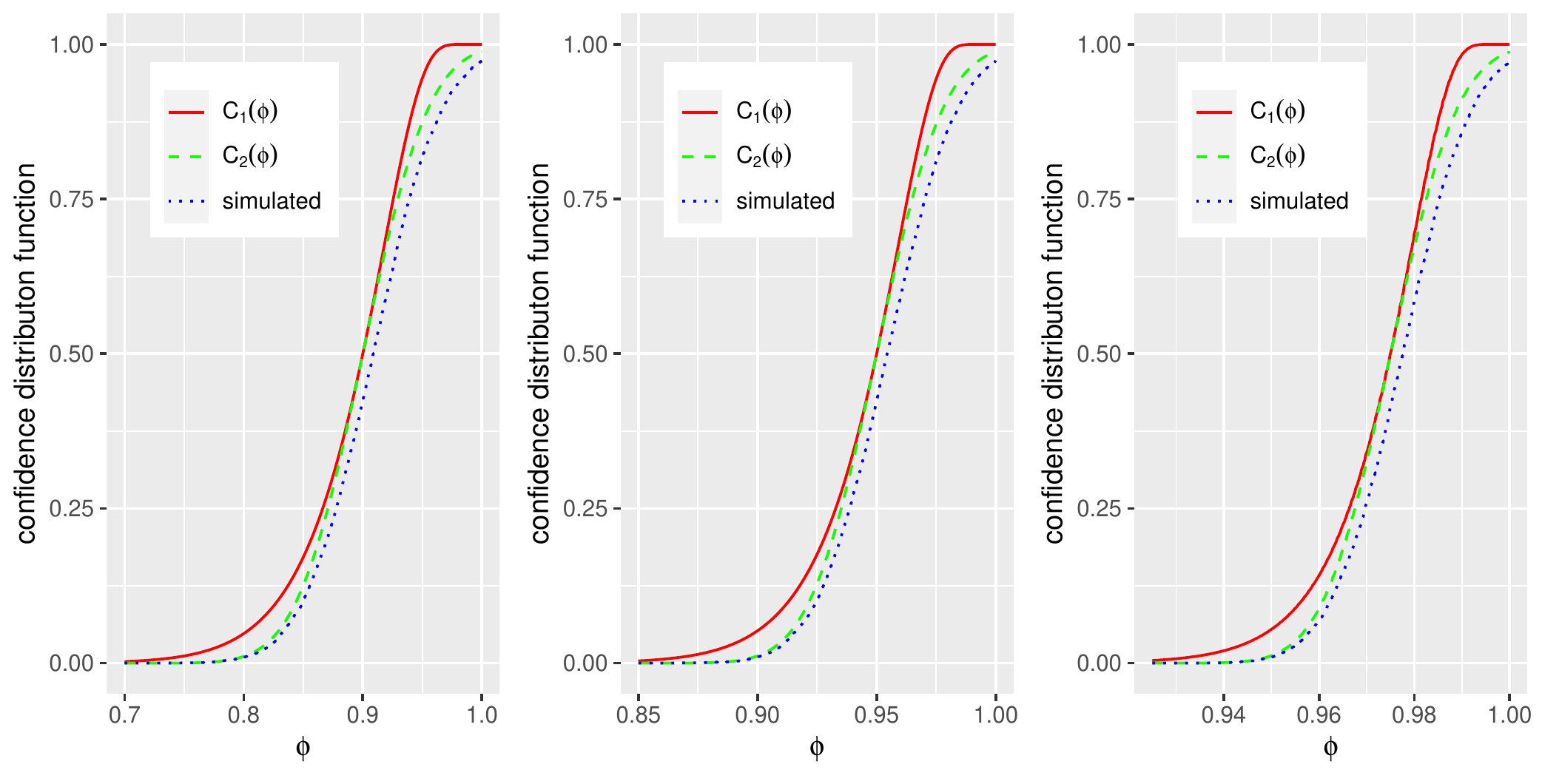}
\end{center}
\caption{Simulated and asymptotic confidence distributions, $n=100, 200, 400$ and $\hat\phi_{obs}=0.90, 0.95, 0.975$ (i.e. $c=10$ in (\ref{phic})) from left to right.}
\end{figure}

\begin{figure}[htb]
\begin{center}
\includegraphics[width=70mm,height=70mm]{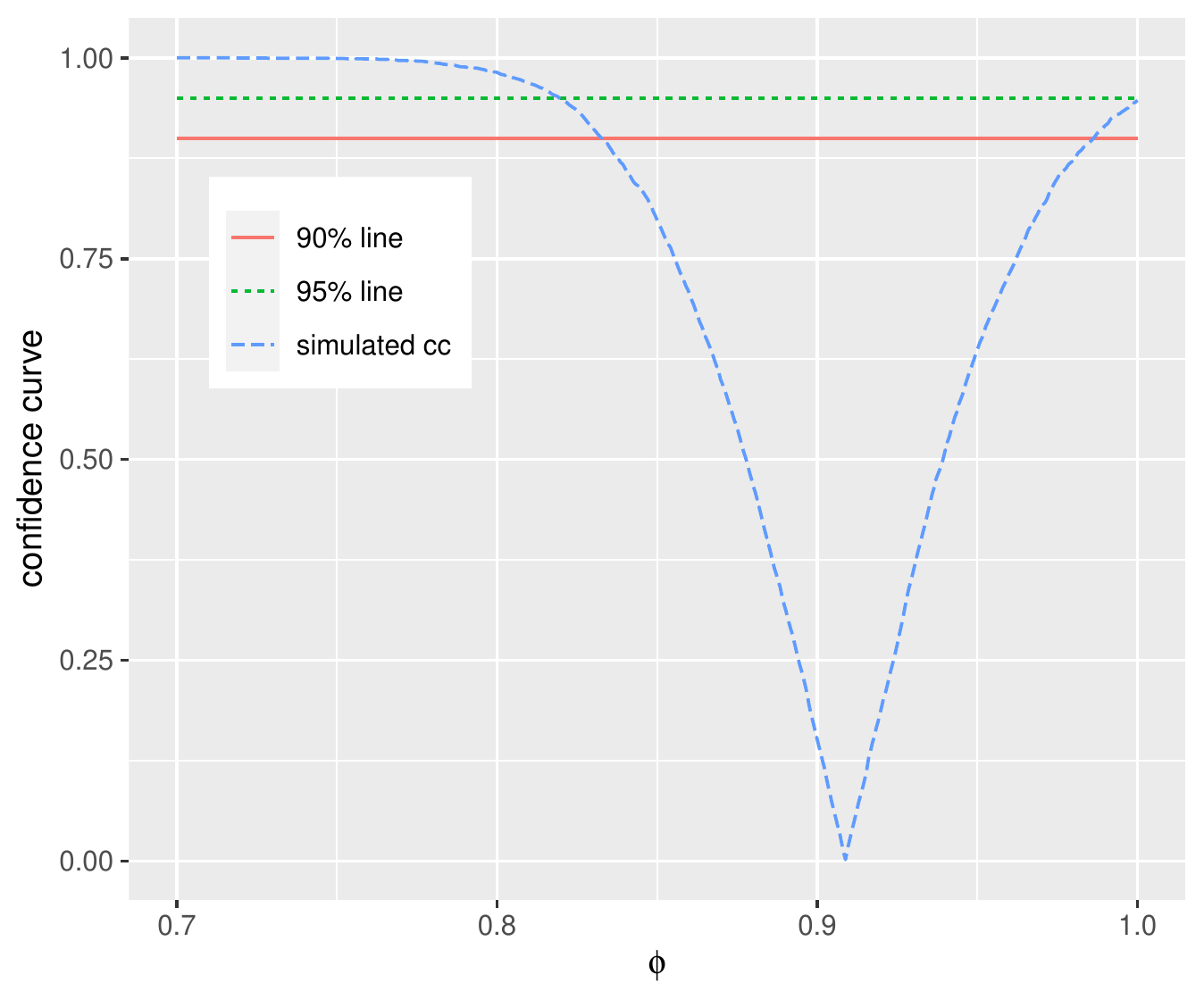}
\end{center}
\caption{Simulated empirical confidence curve, $\hat\phi_{obs}=0.90$, $n=100$.}
\end{figure}

\subsection{Confidence density function}

From (\ref{asarcdf1}), denoting the standard normal density function by $\varphi(\cdot)$, for $|\phi|<1$ we obtain the approximate confidence denstity as
\begin{align}
c_1(\phi)&
=\frac{d}{d\phi}\Phi\left(\sqrt{n}\frac{\phi-\hat\phi_{obs}}{\sqrt{1-\phi^2}}\right)=\frac{d}{d\phi}\left(\sqrt{n}\frac{\phi-\hat\phi_{obs}}{\sqrt{1-\phi^2}}\right)\varphi\left(\sqrt{n}\frac{\phi-\hat\phi_{obs}}{\sqrt{1-\phi^2}}\right)\notag\\
&=\sqrt{n}\frac{1-\hat\phi_{obs}\phi}{(1-\phi^2)^{3/2}}\varphi\left(\sqrt{n}\frac{\phi-\hat\phi_{obs}}{\sqrt{1-\phi^2}}\right)\notag\\
&=\sqrt{n}\frac{1-\hat\phi_{obs}\phi}{(1-\phi^2)^{3/2}}
\frac{1}{\sqrt{2\pi}}\exp\left\{-\frac{n}{2}\frac{(\phi-\hat\phi_{obs})^2}{1-\phi^2}\right\}.\label{asardf1}
\end{align}
Similarly, as in (\ref{asardf1}), the asymptotic approximation in (\ref{asarcdf2}) gives the confidence density
\begin{equation}
c_2(\phi)=\sqrt{n}\frac{1}{\sqrt{1-\hat\phi_{obs}^2}}
\frac{1}{\sqrt{2\pi}}\exp\left\{-\frac{n}{2}\frac{(\phi-\hat\phi_{obs})^2}{1-\hat\phi_{obs}^2}\right\}.\label{asardf2}
\end{equation}

A typical simulation result for large $n$ is given in Figure 4. The confidence density is the derivative of the confidence distribution, but it turns out that plain numerical differentiation gives a very noisy result. For general functions, this type of problem is addressed by Knowles and Renka (2014), whose main idea is to at first construct a smooth approximation of the function to be differentiated. The approximations may be obtained e.g. by fitting polynomials via regressions, or by fitting splines. 

We found the polynomial approach attractive, but because of the properties of distribution functions, it needed to be modified. We decided on a simple solution that guarantees that the approximation can be viewed as a distribution function.
To this end, we approximated the empirical confidence distribution function by $C(\phi)=\Phi(a+b\phi)$,
where $\Phi(\cdot)$ is the standard normal cumulative distribution function, and $a$ and $b$ are some constants. These constants may be estimated by the linear regression $\Phi^{-1}\{C(\phi)\}=a+b\phi$, giving $\hat a$ and $\hat b$. The estimated empirical confidence density is then obtained as the first derivative
$$c_{emp}(\phi)=\hat b\varphi(\hat a+\hat b\phi),$$
where $\varphi(\cdot)$ is the standard normal density function, see further Appendix 1.

\begin{figure}[htb]
\begin{center}
\includegraphics[width=100mm]{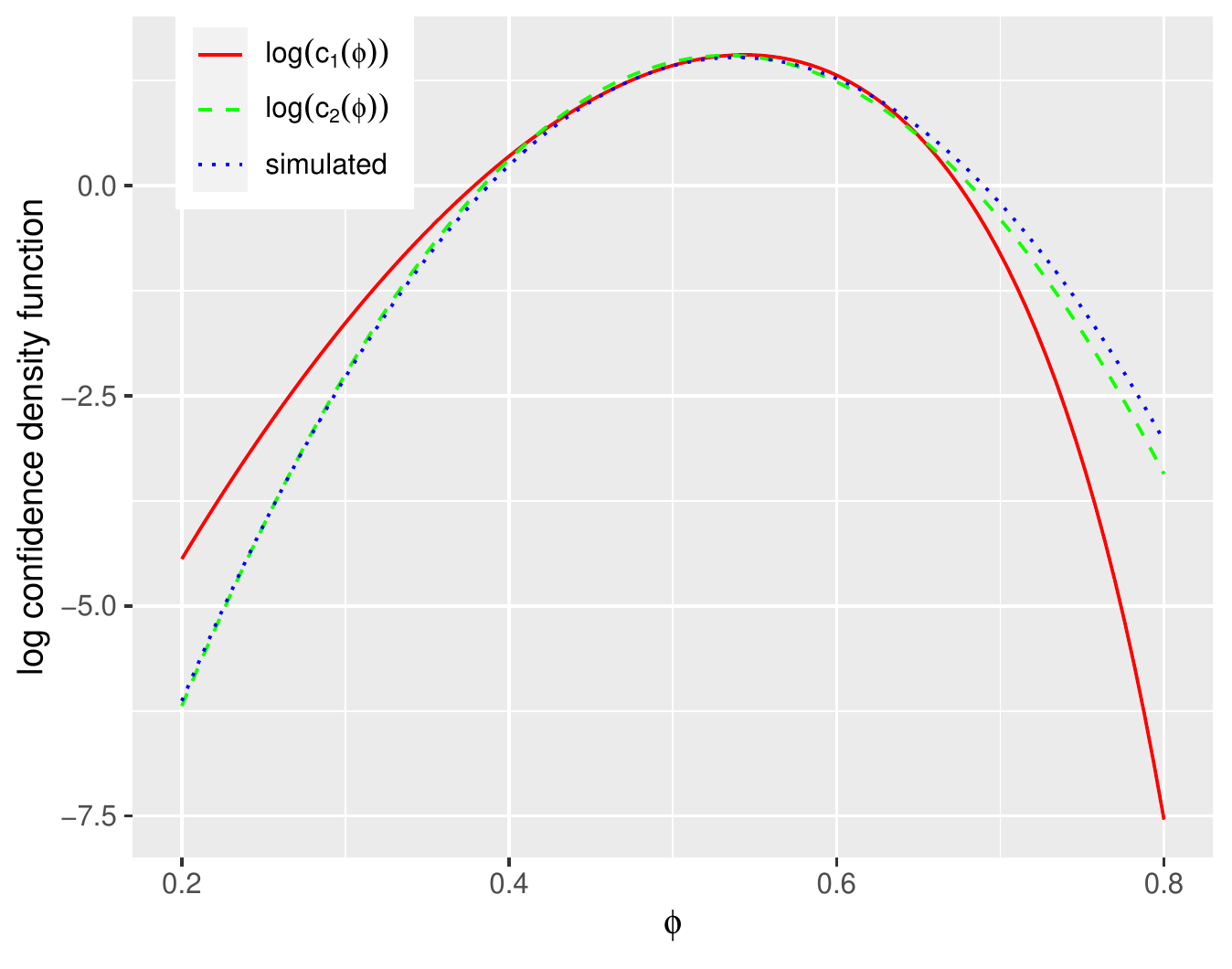}
\end{center}
\caption{Log confidence density functions, for $n=100$, $10^5$ replicates.  In this simulation, $\hat\phi_{obs}=0.533$.
}
\end{figure}

In Figure 4, approximating the (log) empirical confidence density function (in blue), the approximation in (\ref{asardf1}) (red) only works well in the near vicinity of the maximum likelihood estimate $\hat\phi_{obs}$. The alternative approximation of (\ref{asardf2}) (green) works astonishingly well, throughout basically coinciding with the empirical confidence density function.
In the following, the only asymptotic approximation that we will consider is the one in (\ref{asardf2}).

Figure 1 in the appendix gives similar results for a $\hat\phi_{obs}$ close to 0.8.

\subsection{Relating to the likelihood}

Assuming $y_0=0$, the likelihood is
\begin{align}
L(\phi,\sigma^2)&=\frac{1}{\sqrt{2\pi\sigma^2}}\exp\left(-\frac{y_1^2}{2\sigma^2}\right)\prod_{t=2}^n\frac{1}{\sqrt{2\pi\sigma^2}}
\exp\left\{-\frac{(y_t-\phi y_{t-1})^2}{2\sigma^2}\right\}\notag\\
&=(2\pi\sigma^2)^{-n/2}\exp\left(-\frac{1}{2\sigma^2}A\right),
\label{arlik}
\end{align}
where, because (cf (\ref{MLE})) $\sum_{t=1}^ny_{t-1}y_t=\hat\phi_{obs}\sum_{t=1}^ny_{t-1}^2$,
\begin{align}
A&=y_1^2+\sum_{t=2}^n(y_t-\phi y_{t-1})^2
=\sum_{t=1}^n y_t^2-2\phi\sum_{t=1}^ny_{t-1}y_t+\phi^2\sum_{t=1}^ny_{t-1}^2\notag\\
&=y_n^2+\sum_{t=1}^ny_{t-1}^2(1-2\hat\phi_{obs}\phi+\phi^2).
\label{A}
\end{align}
Hence, the approximation in (\ref{asardf2}), (\ref{arlik}) and (\ref{A}) yield
$$\frac{c_2(\phi)}{L(\phi,\sigma^2)}=\sqrt{n}\frac{1}{\sqrt{1-\hat\phi_{obs}^2}}(2\pi)^{(n-1)/2}(\sigma^2)^{n/2}\exp\left(-\frac{1}{2\sigma^2}B\right),$$
where
$$B=n\sigma^2\frac{(\phi-\hat\phi_{obs})^2}{1-\hat\phi_{obs}^2}-y_n^2-\sum_{t=1}^ny_{t-1}^2(1-2\hat\phi_{obs}\phi+\phi^2).$$

In the following, we will use $O_p(n^{-\alpha})$ to denote a term on the form $n^{-\alpha} V$, where $V$ is a non degenerate random variable. Similarly, an $o_p(n^{-\alpha})$ term tends to zero in probability as $n\to\infty$, when multiplied by $n^{\alpha}$. (For a more precise definitions, see Mann and Wald, 1943). Finally, an $O(n^{-\alpha})$ term tends to a nonzero finite constant as $n\to\infty$, when multiplied by $n^{\alpha}$.

We have the following result. 

\begin{proposition}
Say that the data are generated from (\ref{AR1}) with $y_0=0$ and with parameter $\phi_0$, where $|\phi_0|<1$. Then, as $n\to\infty$, for all $|\phi|<1$,
$$n^{-1}\log\left\{\frac{c_2(\phi)}{L(\phi,\sigma^2)}\right\}=\frac{\log(2\pi\sigma^2)+1}{2}+\frac{1}{2}\frac{\log n}{n}+n^{-1/2}h(\phi,\phi_0)U+o_p(n^{-1/2}),$$
where $U$ is a standard normal random variable and
$$h(\phi,\phi_0)=\frac{1-\phi_0-\phi_0^2+\phi(1+\phi_0^2)-\phi^2\phi_0}{(1-\phi_0^2)^{3/2}}.$$
\end{proposition}
\begin{proof}
See Appendix 2.
\end{proof}

In view of the simulations below, note that $(\log(2\pi)+1)/2\approx 1.419$. Also, observe that the expectation of the right-hand side of proposition 1 equals the constant $(\log(2\pi\sigma^2)+1)/2$ plus an error term of order smaller than $n^{-1/2}$, but we may not deduce that the order of the error term is smaller than  $(\log n)/n$.
Moreover, observe that the result of proposition 1 also holds when $\sigma^2$ is replaced by its maximum likelihood estimator,
\begin{equation}
\hat\sigma^2=n^{-1}\sum_{t=2}^n(y_t-\hat\phi_{obs}y_{t-1})^2
=n^{-1}\left\{\sum_{t=2}^ny_t^2-\frac{\left(\sum_{t=2}^n y_{t-1}y_t\right)^2}{\sum_{t=2}^n y_{t-1}^2}\right\},\label{s2}
\end{equation}
which is consistent for $\sigma^2$, see Shumway and Stoffer (2017), p. 125.

In Figures 5 and 6, we have simulated the means (over a large number of replicates) of the log of $c_{emp}(\phi)/L(\phi,\sigma^2)$ divided by $n$, for $n\in\{50,100,200,400\}$, for the cases with known and unknown $\sigma^2$, respectively.  The generating parameters are chosen to be $\phi_0=0.5$, $\sigma^2=1$. Relative to the scales on the y axes, we may note that the curves are almost flat, and also that they become flatter the higher the sample size. This is an empirical indication that with a uniform prior on $\phi$, for large $n$ the posterior is approximately equal to the confidence density, in the stationary case $\phi\in(-1,1)$. Moreover, note the convergence rate that seems to be almost $n^{-1}$. Compare to proposition 1, which contains a deterministic error term of order $(\log n)/n$, while the stochastic error term has expectation zero. In our simulation, we evaluate the mean over many replicates, corresponding to the expectation.
We have also performed simulations for $\phi_0=0.8$, see the appendix. The results are similar.

\begin{figure}[htb]
\begin{center}
\includegraphics[width=100mm]{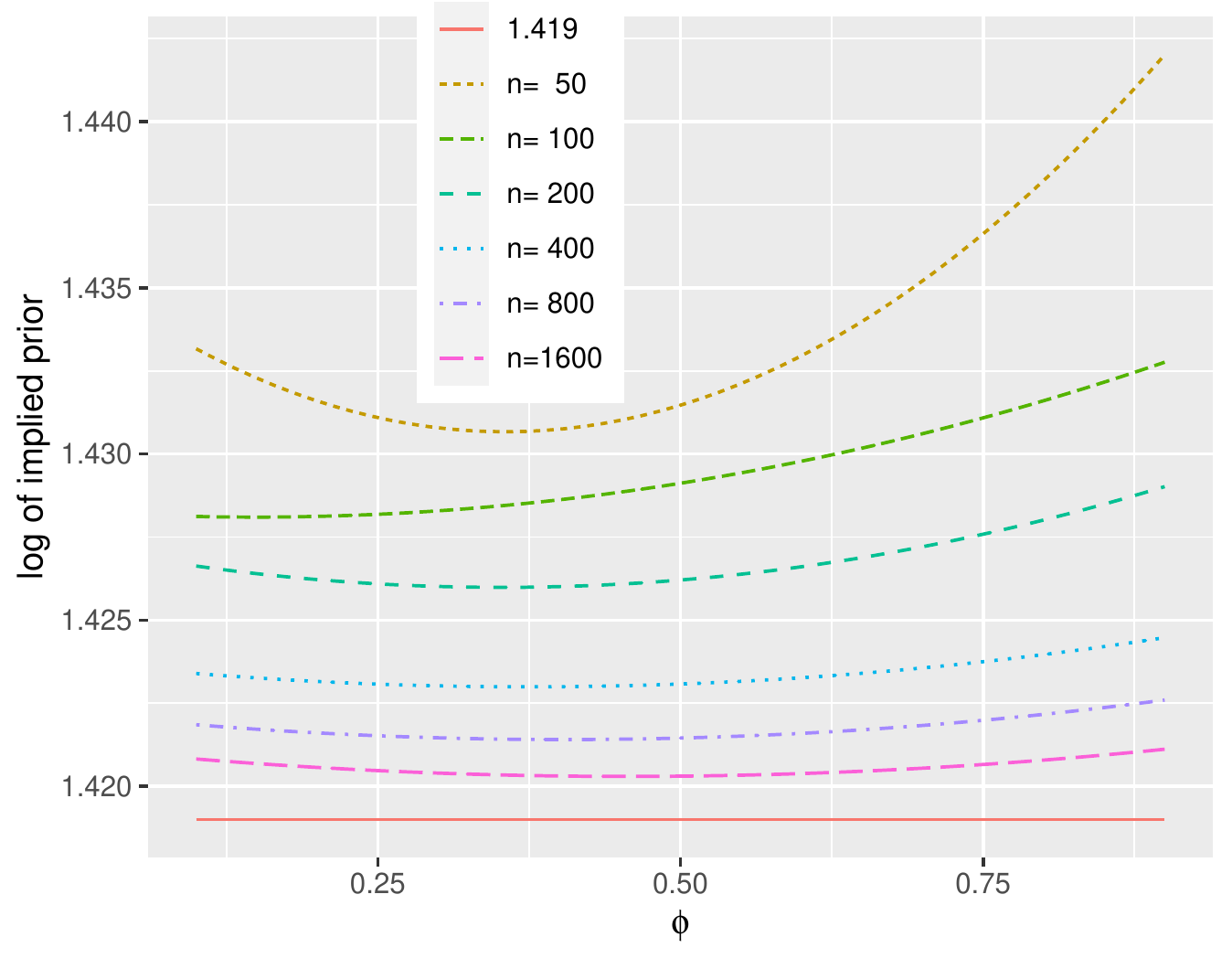}
\end{center}
\caption{Simulated logs of implied priors $c_{emp}(\phi)/L(\phi,\sigma^2)$ for generating parameters $\phi_0=0.5$, $\sigma^2=1$, known $\sigma^2$. We have 1000 replicates for each $\hat\phi_{obs}$, and 50 000 replicates of $\hat\phi_{obs}$.
}
\end{figure}

\begin{figure}[htb]
\begin{center}
\includegraphics[width=100mm]{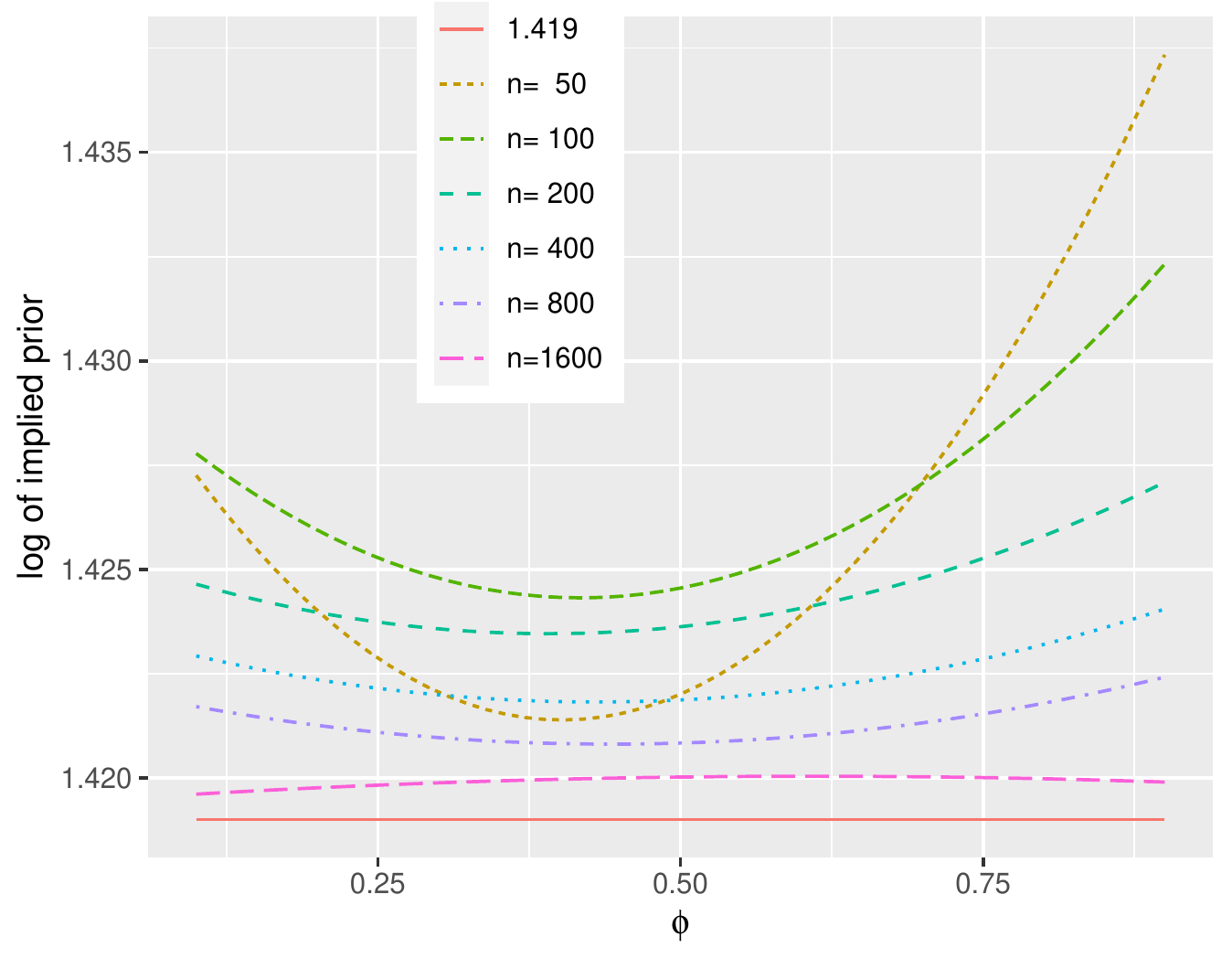}
\end{center}
\caption{Simulated logs of implied priors $c_{emp}(\phi)/L(\phi,\sigma^2)$ for generating parameters $\phi_0=0.5$, $\sigma^2=1$, unknown $\sigma^2$. We have 1000 replicates for each $\hat\phi_{obs}$, and 50 000 replicates of $\hat\phi_{obs}$.
}
\end{figure}

In a Bayesian analysis more fair to the problem, the flat prior should be complemented by a spike at one. In practice, it is hard to know how large it should be, but for the sake of argument, let us add a spike with height $b$ of the prior implied confidence curve at $\phi=1$. This means that for $-1<\phi< 1$, the confidence density is (cf (\ref{c}))
\begin{equation}
c(\phi)=L(\phi;y)h(\phi)=aL(\phi;y),\label{cphi}
\end{equation}
where $L(\phi;y)$ is the likelihood and $a$ is a constant that we want to determine, but for the confidence distribution $C(\phi)$, we should have $C(1)=1-b$. Hence, it follows from (\ref{cphi}) that
$$1-b=C(1)=a\int_{-1}^1L(\phi;y)d\phi,$$
i.e.
$$a=\frac{1-b}{\int_{-1}^1L(\phi;y)d\phi}.$$
Conclusively, (\ref{cphi}) yields that the confidence distribution function becomes
\begin{equation}
C(\phi)=a\int_{-1}^\phi L(\phi^*;y)d\phi^*
=(1-b)\frac{\int_{-1}^\phi L(\phi^*;y)d\phi^*}{\int_{-1}^1 L(\phi^*;y)d\phi^*},\label{Cnorm}
\end{equation}
where $\int_{-1}^\phi L(\phi^*;y)d\phi^*$ is the integrated likelihood at $\phi$. 

\section{Empirical examples}

In this section, we present two empirical examples. The first one is the log exchange rate between the Swedish Krona and Euro with daily data from July 1 2021 until the end of the year, giving 132 days with observations. The data was downloaded from ECB (2022). The series is depicted in Figure 7.

\begin{figure}[htb]
\begin{center}
\includegraphics[width=100mm]{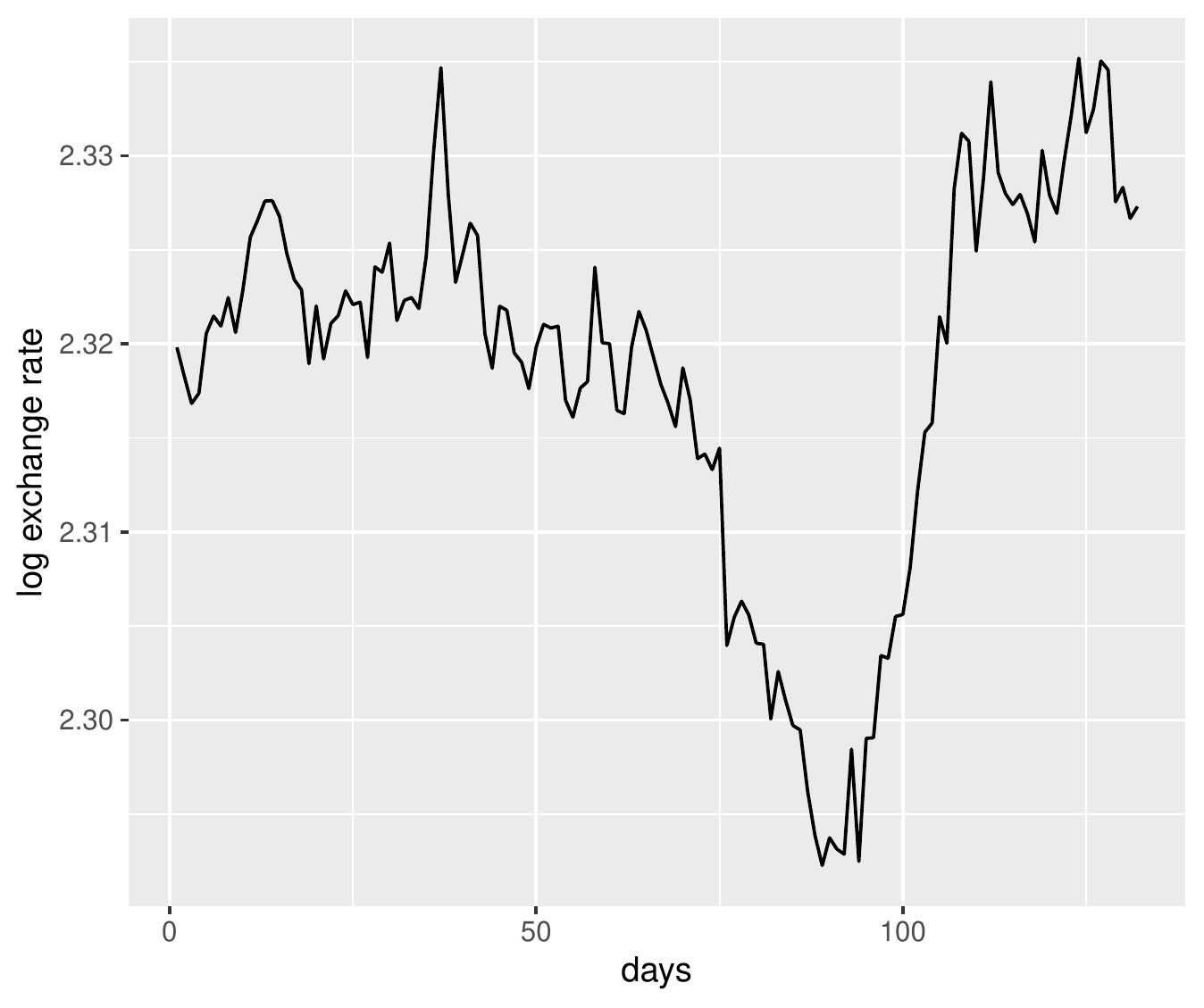}
\end{center}
\caption{Log exchange rate SKr/Euro, second half of 2021.
}
\end{figure}

After subtracting the mean, we fitted an AR(1) model to the data and obtained the maximum likelihood estimate $\hat\phi_{obs}=0.963$ by using (\ref{MLE}). The confidence distribution is obtained by bootstrap, for each $\phi$ by resampling the residuals and then generating the bootstrap sample based on these. For stability of results, the resampling scheme is the same for all $\phi$, see further Appendix 1. Throughout, we used $10^6$ bootstrap replications.

Going directly to the empirical confidence curve, it is given in Figure 8. We may note that the confidence curve is about 0.76 at $\phi=1$, which means that the p value of a corresponding unit root test is about $(1-0.76)/2=0.12$. The 95\% and 90\% confidence intervals for $\phi$ can be read off as $(0.918,\ 1)$ and $(0.926,\ 1)$, respectively.

\begin{figure}[htb]
\begin{center}
\includegraphics[width=100mm]{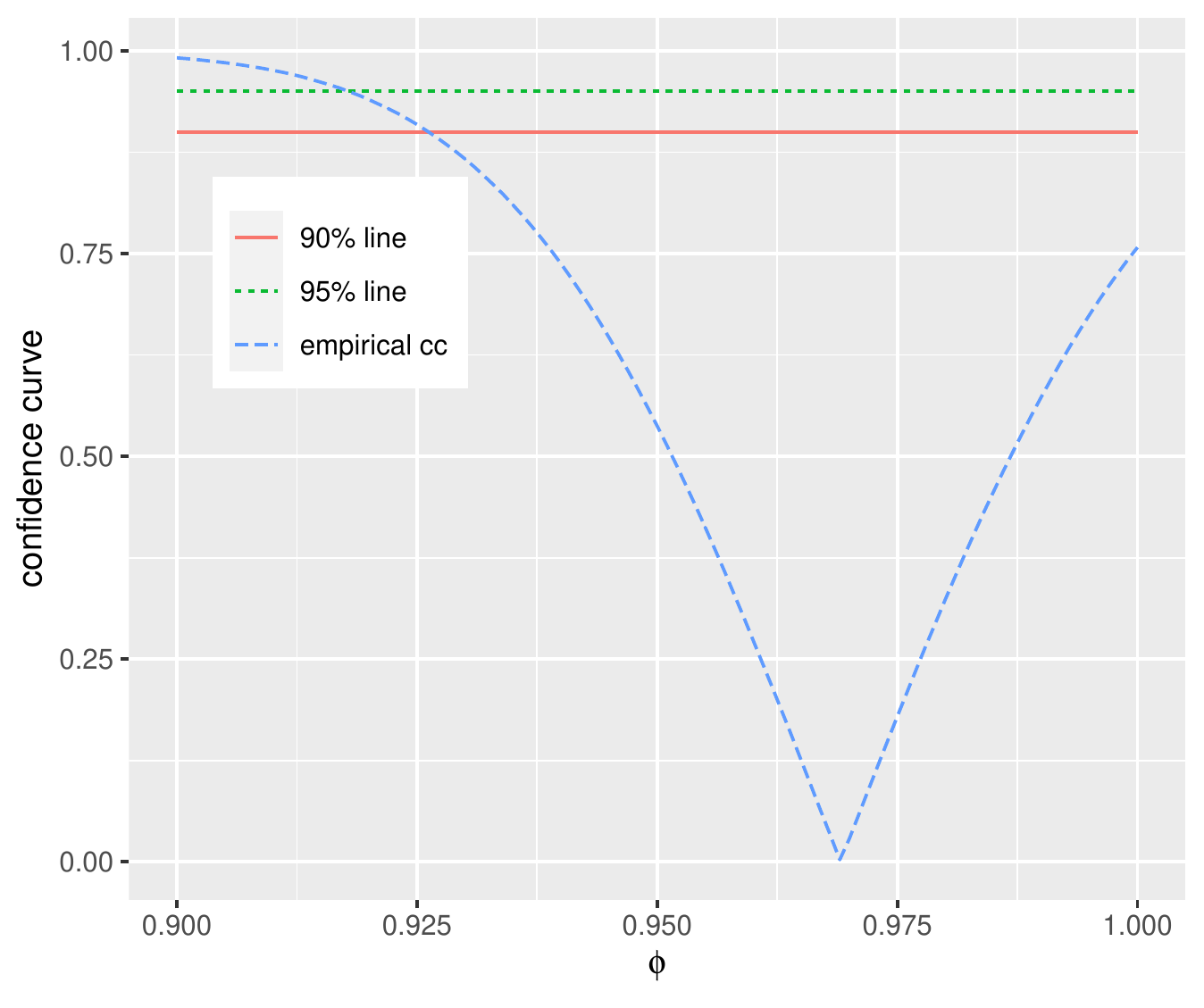}
\end{center}
\caption{Log exchange rate SKr/Euro, confidence curve.
}
\end{figure} 

Consider for a moment that we restrict the parameter space to $\phi\in(-1,1)$. Then, according to proposition 1, for large samples the implied non informative prior is close to constant. With such a prior, the confidence density is proportional to the likelihood (cf (\ref{c})). The likelihood is obtained from (\ref{arlik}) with $\hat\sigma^2$ from (\ref{s2}) inserted for $\sigma^2$. Thus, strictly speaking, this is a profile likelihood. This also means that what we do is not a proper Bayesian analysis, since we avoid putting a prior on $\sigma^2$. Because of the proportionality, we may obtain the corresponding confidence distribution function from the integral
$$L_1(\phi)=\int_{-1}^\phi L(\phi^*)d\phi^*,$$
divided by $L_1(1)$ to make it a proper distribution function. From that, we may in turn calculate the corresponding confidence curve (cf (\ref{cc}) and Appendix 1). In a Bayesian setting, this confidence curve gives the equally tailed credible intervals for $\phi$ in exactly the same way as an ordinary confidence curve gives confidence intervals. For our data, Figure 9 gives this confidence curve. 
The 95\% and 90\% credible intervals for $\phi$ can be read off as $(0.916,\ 0.996)$ and $(0.923,\ 0.993)$, respectively. It is natural that the flat prior on $\phi\in(-1,1)$ results in excluding 1 from all credible intervals.

\begin{figure}[htb]
\begin{center}
\includegraphics[width=100mm]{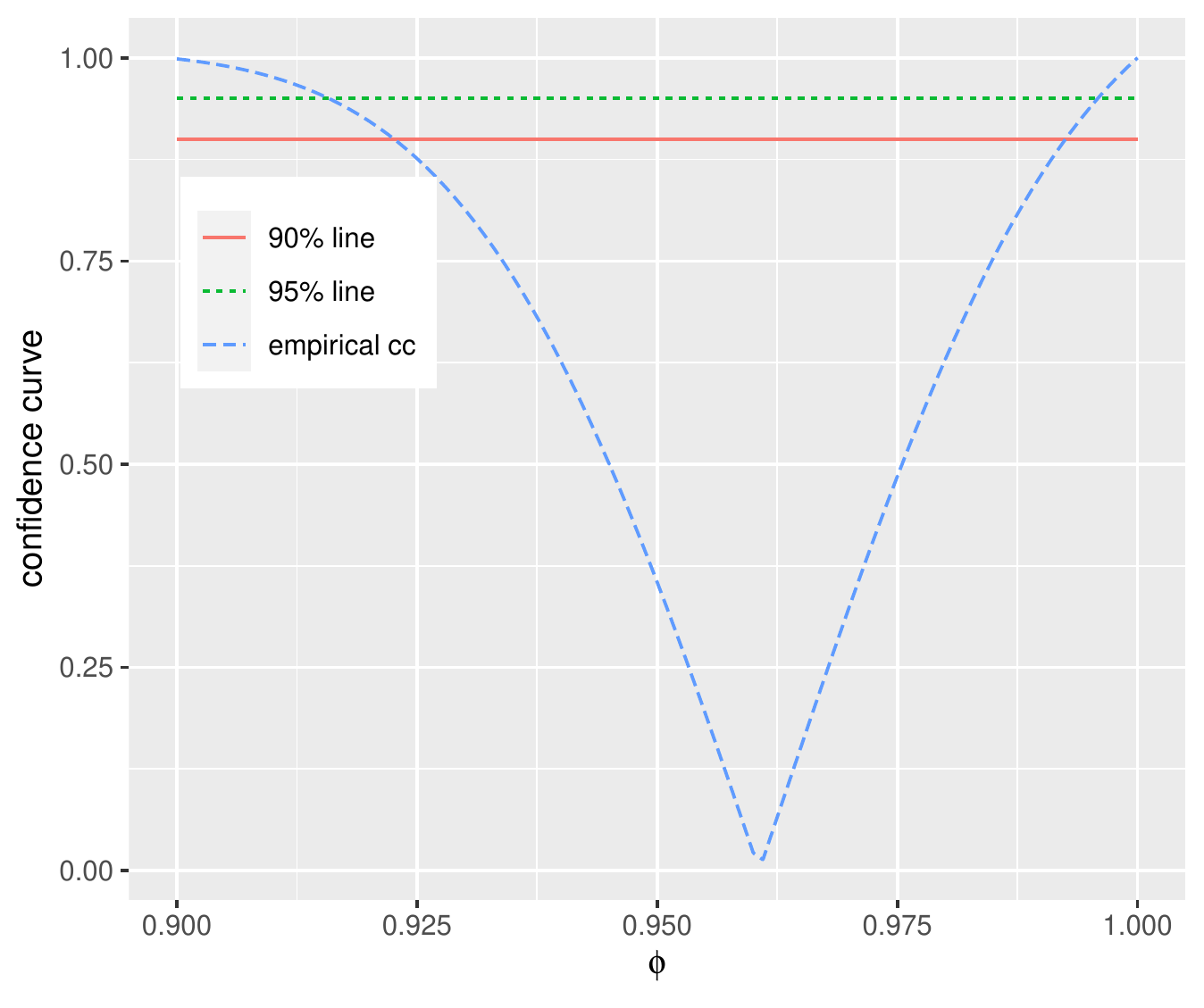}
\end{center}
\caption{Log exchange rate SKr/Euro, `confidence curve' obtained via a flat Bayesian prior.
}
\end{figure}

If we try to add a spike at one in the prior, how large should it be? A pure Bayesian would never look at the data to specify the prior. But if we do this anyway, we can find from the numbers behind Figure 8 that we need a spike of height $1-0.76=0.24$. Now, we apply (\ref{Cnorm}) with $b=0.24$ to obtain Figure 10 (for details, see Appendix 1).
There, we plot the so obtained confidence curve together with the one from the frequentist analysis also given in Figure 8. 

We may note that the Bayesian analysis somewhat shifts the curve to the left, but not with much. Based on the new confidence curve, the 95\% and 90\% credible intervals for $\phi$ are $(0.917,\ 1)$ and $(0.924,\ 1)$, respectively, i.e. very close to the frequentist confidence intervals.

\begin{figure}[htb]
\begin{center}
\includegraphics[width=100mm]{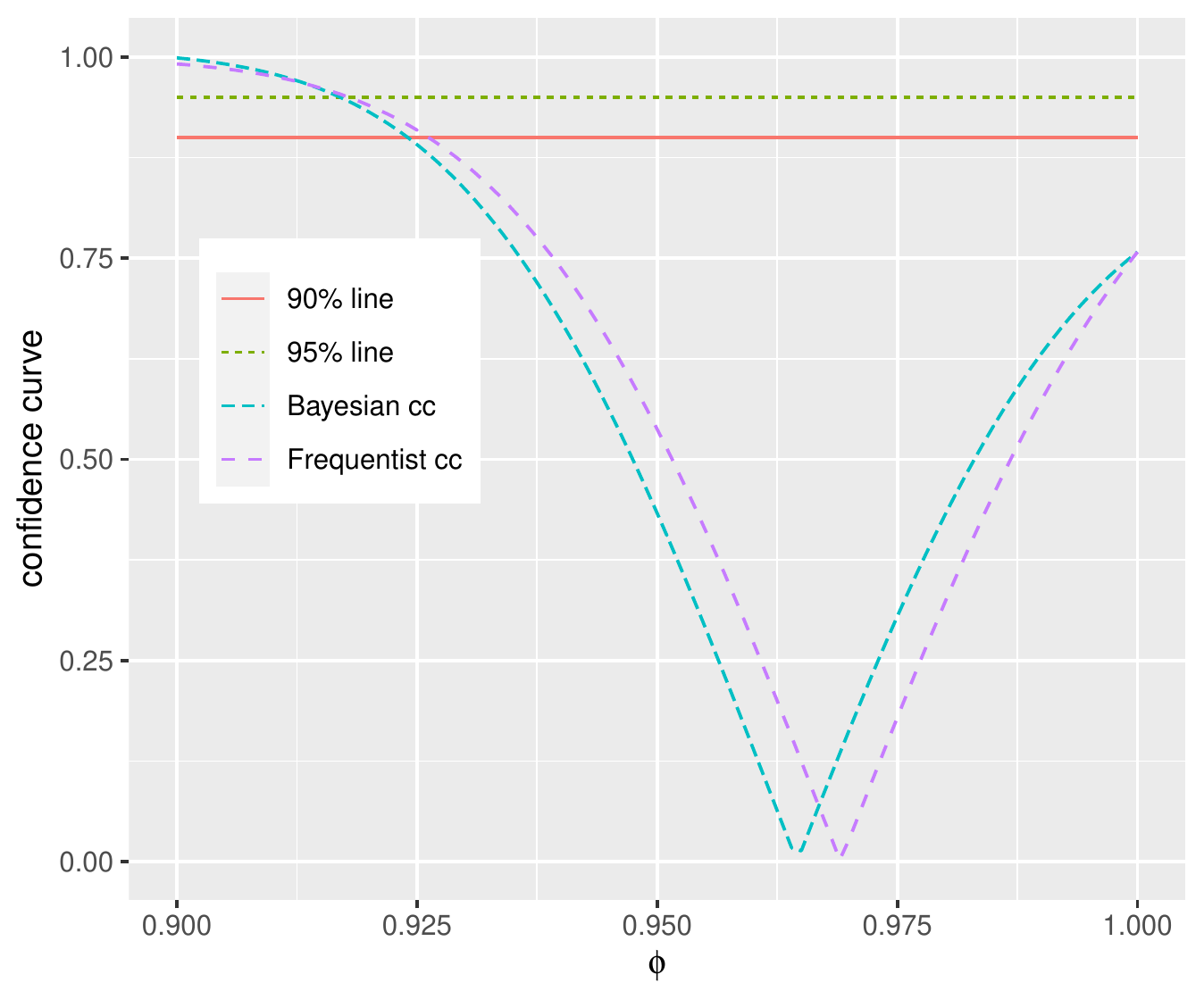}
\end{center}
\caption{Log exchange rate SKr/Euro, Bayesian and frequentist confidence curves.}
\end{figure}

Our second example is the unemployment rate in the US for citizens aged 25-54, yearly data for 1948-2020. The data was collected from gapminder (2022). In this case, after subtracting the mean, the maximum likelihood estimate from (\ref{MLE}) is $\hat\phi_{obs}=0.732$, and a unit root is firmly rejected. The data is plotted in Figure 11.
\begin{figure}[htb]
\begin{center}
\includegraphics[width=100mm]{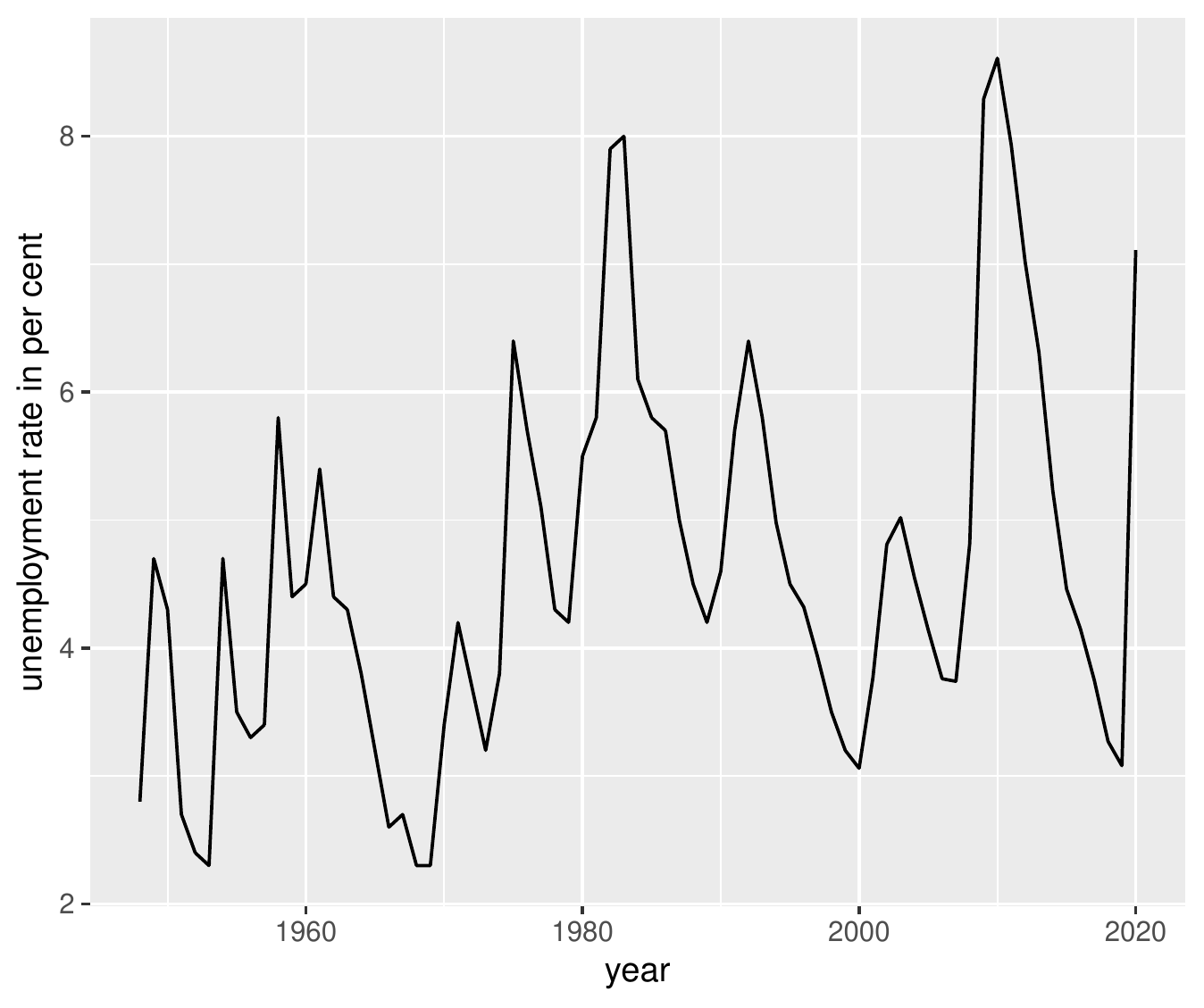}
\end{center}
\caption{US unemployment rate, ages 25-54, years 1948-2020.}
\end{figure}
An analysis on the same lines as above gives confidence curves as in Figure 12. Again, the Bayesian curve is slightly to the left of the frequentist curve. The frequentist 95\% and 90\% confidence intervals are $(0.578,\ 0.902)$ and $(0.606,\ 0.875)$, respectively, while the corresponding Bayesian credible intervals are $(0.573,\ 0.891)$ and $(0.598,\ 0.865)$.

\begin{figure}[htb]
\begin{center}
\includegraphics[width=100mm]{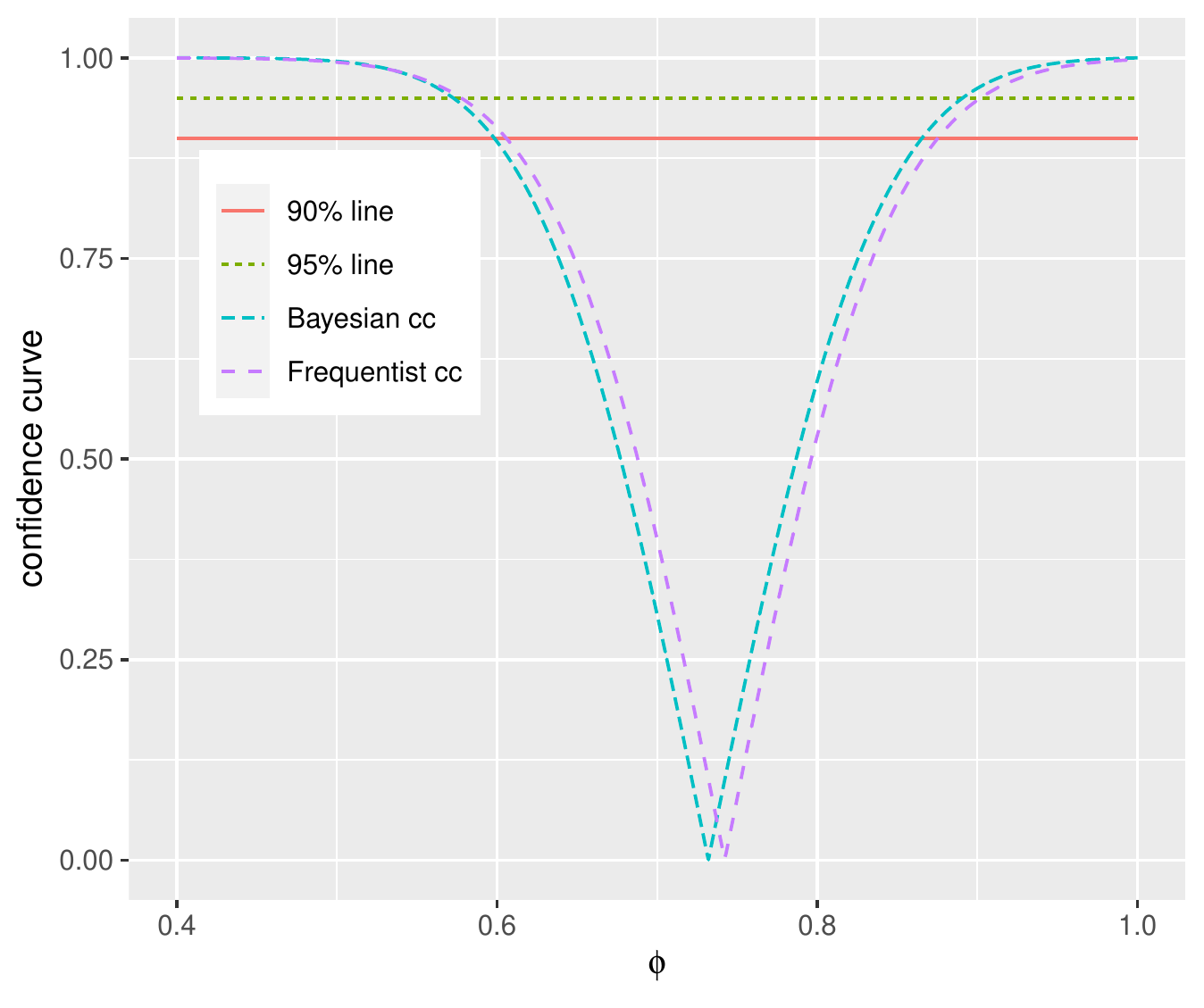}
\end{center}
\caption{US unemployment rate, Bayesian and frequentist confidence curves.}
\end{figure}

\section{Concluding remarks}

The paper recalls the notion of confidence distributions and applies it to inference about the parameter in an autoregressive process of order one with no intercept. At first, we investigate two ways of performing normal approximation in the stationary case. Simulations indicate that the version where the estimated parameter is plugged in into the variance of the estimator is the one that works best. Moreover, in the presence of a unit  parameter (a unit root), we illustrate that the empirical confidence distribution does not reach the value 1. 

Next, the focus is on non informative Bayesian analysis, where we show that in the presence of a large sample, the prior has to be flat in the stationary case. When the true autoregressive coefficient  parameter value is one, the prior distribution has to contain a spike at that value. Finally, we give two empirical examples, one where a unit root can not be rejected and one where it can. In the latter case, the flat prior seems to work well.

The present paper demonstrates some of the powers of the confidence distribution paradigm. Still, one might want to go for Bayesian analysis because of its many benefits like coherence, interpretation and so forth. But the question on how to formulate the prior always remains. If no prior knowledge about a parameter is at hand, it is tempting to use a non informative one. Inducing a non informative prior by confidence methods presents one way to do so.  Our paper illustrates how this can be done in cases where the parameter of interest might be located at the edge of the parameter space. Hence, confidence methods might constitute a welcome addition to the Bayesian toolbox.

Future work could concern generalizations of the simple autoregressive model to higher lag orders. Another interesting scope is to look at other cases where the parameter space is restricted, like testing independence in a dependent Poisson model, see e.g. Larsson (2022).

\section*{Supplemental material}

Three supplemental appendicies are given. Appendix 1 gives details on the simulations and numerical calculations that lead to the figures.
In appendix 2, a proof of proposition 1 is found. Finally, appendix 3 provides some extra figures with simulation results, and comments on these figures.

\section*{Acknowledgements}

I am grateful to Svante Janson and Shaobo Jin for helpful discussions and comments.

\section*{References}
Birnbaum, A. (1961) "Confidence curves: An omnibus technique for estimation and testing statistical hypotheses," \emph{Journal of the American Statistical Association}, 56, 246-249.
\\[1.4mm]
Cox, D. R. (1958) "Some problems connected with statistical inference," \emph{Annals of Mathematical Statistics}, 29, 357-372.
\\[1.4mm]
ECB (2022) https://www.ecb.europa.eu/stats/policy and exchange rates/euro reference exchange rates/html/sek.xml, downloaded August 30, 2022.
\\[1.4mm]
Elliott, G., and Stock, J.H. (2001) "Confidence intervals for autoregressive coefficients near one," \emph{Journal of Econometrics}, 103, 155-181.
\\[1.4mm]
Fisher, R. A. (1930) "Inverse probability," \emph{Proceedings of the Cambridge Philosophical Society}, 26, 528-535.
\\[1.4mm]
Hald, A. (2007) \emph{A History of Parametric Statistical Inference from Bernoulli to Fisher, 1713-1935}, New York: Springer.
\\[1.4mm]
Gapminder (2022) https://www.gapminder.org/data/, downloaded September 8, 2022.
\\[1.4mm]
Gnedenko, B. V. (1989) \emph{The Theory of Probability, 4th ed.}, New York: Chelsea Publishing Company.
\\[1.4mm]
Jeffreys, H. (1931) \emph{Theory of Probability}, Cambridge: Cambridge University Press.
\\[1.4mm]
Jeffreys, H. (1946) "An invariant form for the prior probability in estimation problems," \emph{Proceedings of the Royal Statistical Society, Series A}, 186, 453-461. 
\\[1.4mm]
Knowles, I., and Renka, R. J.  (2014) "Methods for numerical differentiation of noisy data," \emph{Electronic Journal of Differential Equations, Conference 21}, 235-246.
\\[1.4mm]
Larsson, R. (2022) "Bartlett correction of an independence test in a multivariate Poisson model," \emph{Statistica Neerlandica}, 76, 391-417.
\\[1.4mm]
Lindley, D. V. (1958) "Fiducial distributions and Bayes' Theorem," \emph{Journal of the Royal Statistical Society, Series B}, 20, 102-107.
\\[1.4mm]
Mann, H. B., and Wald, A.  (1943) "On stochastic limit and order relationships," \emph{Annals of Mathematical Statistics}, 14, 217-226.
\\[1.4mm]
Neyman, J. (1941) "Fiducial argument and the theory of confidence intervals," \emph{Biometrika}, 32, 128-150.
\\[1.4mm]
Perron, P. (1989) "The calculation of the limiting distribution of the least-squares estimator in a near-integrated model," \emph{Econometric Theory}, 5, 241-255.
\\[1.4mm]
Phillips, P. C. B. (1987) "Towards a unified asymptotic theory for autoregression," \emph{Biometrika}, 74, 535-547.
\\[1.4mm]
Phillips, P. C. B. (2014) "On confidence intervals for autoregressive roots and predictive regression," \emph{Econometrica}, 82, 177-1195.
\\[1.4mm]
Schweder, T., and Hjort, N. L. (2016) \emph{Confidence, Likelihood, Probability - Statistical Inference with Confidence Distributions}, Cambridge University Press: New York.
\\[1.4mm]
Shumway, R. H., Stoffer, D. S. (2017) \emph{Time Series Analysis and its Applications -- With R Examples. 4th edition},
New York: Springer.
\\[1.4mm]
Xie, M., and Singh, K. (2013) "Confidence distribution, the frequentist distribution estimator of a parameter: a review," \emph{International Statistical review}, 81, 3-39.
\\[1.4mm]

\end{document}


\title{Confidence distributions for the autoregressive parameter, supplemental material}
\date{}
\maketitle

\section*{Appendix 1: Simulation setups}

In this Appendix, we shortly describe the numerical calculations and simulations that lie behind the figures of the paper. 

\subsection*{Figures 1-3: Confidence distributions}

The simulation to calculate confidence distributions, that gives Figures 1-3, is as follows.

\begin{enumerate}
\item The observed MLE $\hat\phi_{obs}$ is specified beforehand.
\item Put $h=(1-\phi_{min})/n_\phi$, where $n_\phi$ is some large number. Then, for each $\phi\in\{\phi_{min},\phi_{min}+h,\phi_{min}+2h,...,1\}$, calculate the simulated (empirical) confidence distribution function  $C(\phi)$ as follows:
\begin{enumerate}
\item For each simulation replicate $r=1,...,N$ for some large $N$ do
\begin{enumerate}
\item Given some large $n$, generate independent standard normal random  random variables $\varepsilon_1,...,\varepsilon_n$. 
\item Put $y_1=\varepsilon_1$.
\item For $t=2,...,n$, put $y_t=\phi y_{t-1}+\varepsilon_t$.
\item Calculate the MLE $\hat\phi$ from the generated sample $y_1,...,y_n$.
\item Check if $\hat\phi>\hat\phi_{obs}$.
\end{enumerate}
\item Estimate $C(\phi)$ as the proportion of replicates where $\hat\phi>\hat\phi_{obs}$ (cf (8)).
\item For all $\phi\in\{\phi_{min},\phi_{min}+h,\phi_{min}+2h,...,1\}$, calculate the asymptotic approximations (9) and (10).
\end{enumerate}
\end{enumerate}
To get Figure 3, use $cc(\phi)=|1-2C(\phi)|$, where $C(\phi)$ is the simulated confidence distribution function from above.

\subsection*{Figures 4-6: Confidence densities}

To get the confidence densities in Figure 4, the procedure is as follows.

\begin{enumerate}
\item Fix the `true' parameter value $\phi=\phi_0$. (For Figure 4, we have $\phi_0=0.5$.)
\item Given some large $n$, generate independent standard normal random  random variables $\varepsilon_1,...,\varepsilon_n$. 
\item Put $x_1=\varepsilon_1$.
\item For $t=2,...,n$, put $x_t=\phi_0 x_{t-1}+\varepsilon_t$.
\item Calculate the observed MLE, $\hat\phi_{obs}$, from the generated sample $x_1,...,x_n$.
\item Put $h=(\phi_{max}-\phi_{min})/n_\phi$, where $n_\phi$ is some large number. Then, for each $\phi\in\{\phi_{min},\phi_{min}+h,\phi_{min}+2h,...,\phi_{max}\}$, calculate the simulated (empirical) confidence distribution function  $C(\phi)$ as follows (this is the same as for Figures 1-3):
\begin{enumerate}
\item Given some large $n$, generate independent standard normal random  random variables $\varepsilon_1,...,\varepsilon_n$. 
\item Put $y_1=\varepsilon_1$.
\item For $t=2,...,n$, put $y_t=\phi y_{t-1}+\varepsilon_t$.
\item Calculate the MLE $\hat\phi$ from the generated sample $y_1,...,y_n$.
\item Check if $\hat\phi>\hat\phi_{obs}$ (cf (8)).
\end{enumerate}
\item Estimate $C(\phi)$ as the proportion of replicates where $\hat\phi>\hat\phi_{obs}$.
\item To avoid numerical problems, select all $\phi\in\{\phi_{min},\phi_{min}+h,\phi_{min}+2h,...,\phi_{max}\}$ that are such that $0.01<C(\phi)<0.99$. 
\item For these $\phi$, calculate $\Phi^{-1}\{C(\phi)\}$ and regress on $\phi$ to get the least squares estimates $\hat a$ and $\hat b$ and the `smoothed' estimated confidence distribution function $C_s(\phi)=\Phi(\hat a+\hat b\phi)$. This gives the estimated confidence density as
$c_{emp}(\phi)=C_s'(\phi)=\hat b\varphi(\hat a+\hat b\phi)$.
\item The asymptotic confidence densities are obtained by inserting $\hat\phi_{obs}$ and all selected $\phi$ into (15) and (16).
\end{enumerate}

The confidence densities used for the implied priors of Figures 5 and 6 are obtained in the same way. To get the implied priors, we add calculation of the log of (18) with the generated sample from step 5 above, $x_1,...,x_n$, and $\hat\phi_{obs}$ inserted. For Figure 6, the MLE of $\sigma^2$ from (22) is used.

\subsection*{Figures 8-10:  Bootstrap}

To get the confidence curve in Figure 8, the procedure is as follows.

\begin{enumerate}
\item Input is the observed series $y_1,y_2,...,y_n$, for which we calculate the MLE of the AR coefficient $\hat\phi_{obs}$ from (7).
\item Generate an $n\times N$ matrix of bootstrap indices from the discrete uniform distribution on $1,2,...,n$, where $N$ is the number of bootstrap replicates below. (This is done once and for all, so that for stability, we get the same bootstrap scheme for all $\phi$.)
\item Put $h=(1-\phi_{min})/n_\phi$, where $n_\phi$ is some large number. 
\item For a large number $N$ of bootstrap replicates, and for all $\phi\in\{\phi_{min},\phi_{min}+h,\phi_{min}+2h,...,1\}$:
\begin{enumerate}
\item Resample the `residuals' $e_t=y_t-\phi y_{t-1}$ according to the bootstrap index matrix.
\item Generate a new series $\{y_t^*\}$, where $y_1^*=e_1$, and for $t=2,...,n$, $y_t^*=\phi y_{t-1}^*+e_t$.
\item Calculate the MLE $\hat\phi$ from this series and check if $\hat\phi>\hat\phi_{obs}$.
\end{enumerate}
\item For all $\phi$, estimate $C(\phi)$ as the proportion of bootstrap replicates where $\hat\phi>\hat\phi_{obs}$.
\item Calculate the bootstrap confidence curve through $cc(\phi)=|1-2C(\phi)|$.
\end{enumerate}

For Figure 9, we treat the likelihood, normed by the likelihood at $\phi=1$ as described in section 4, as a confidence distribution function, and then transform it to a confidence curve in the usual way.

The Bayesian confidence curve in Figure 10 is based on the modified confidence distribution function obtained by multiplying the previous Bayesian confidence distribution function with $C(1)$ from Figure 8.

\section*{Appendix 2: Proof of proposition 1}
With $B$ as in (21) in the main text, we will show that $n^{-1}B$ converges to $-\sigma^2$, from which the result easily follows.

By e.g. Shumway and Stoffer (2017), p. 125, we have that as $n\to\infty$, denoting convergence in distribution by $\dlim$,  
$$\sqrt{n}(\hat\phi_{obs}-\phi_0)\dlim U,$$
for a normally distributed variate $U$ with expectation zero and variance $1-\phi_0^2$. We write this as
\begin{equation}
\hat\phi_{obs}=\phi_0+n^{-1/2}U+o_p(n^{-1/2}).\label{phihat}
\end{equation}
Now, (\ref{phihat}) implies
\begin{align}
&(\hat\phi_{obs}-\phi)^2\notag\\
&=\left\{(\hat\phi_{obs}-\phi_0)+(\phi_0-\phi)\right\}^2
=\left\{n^{-1/2}U+o_p(n^{-1/2})+(\phi_0-\phi)\right\}^2\notag\\
&=(\phi_0-\phi)^2+2n^{-1/2}(\phi_0-\phi)U+o_p(n^{-1/2}).
\label{phiobsphi}
\end{align}
Similarly, we have
\begin{align}
&1-2\hat\phi_{obs}\phi+\phi^2\notag\\
&=1-2\left\{(\hat\phi_{obs}-\phi_0)+\phi_0\right\}\phi+\phi^2\notag\\
&=1-2\left(n^{-1/2}U+o_p(n^{-1/2})+\phi_0\right)\phi+\phi^2\notag\\
&=1-2\phi_0\phi+\phi^2-2n^{-1/2}U+o_p(n^{-1/2})\notag\\
&=(\phi_0-\phi)^2+1-\phi_0^2-2n^{-1/2}U+o_p(n^{-1/2}).\label{phiobsphi2}
\end{align}
Moreover, by recursion,
\begin{equation}
y_t=\phi_0 y_{t-1}+\varepsilon_t=\phi_0(\phi_0 y_{t-2}+\varepsilon_{t-1})+\varepsilon_t=...=\sum_{i=0}^{t-1}\phi_0^i\varepsilon_{t-i}.
\label{yt}
\end{equation}
Consequently, for each $t$, $y_t$ is normal with expectation zero and variance $(1-\phi_0^{2t})/(1-\phi_0^2)$. This means that as $n\to\infty$, $(1-\phi_0^2)y_n^2$ is asymptotically $\chi^2$ with one degree of freedom, i.e. $y_n^2=O_p(1)$. 

Furthermore, rewriting (\ref{yt}) as
$$y_t=\sum_{j=1}^t\phi_0^{t-j}\varepsilon_j,$$
we obtain ($i\vee j$ is the maximum of $i$ and $j$)
\begin{align*}
&\sum_{t=1}^n y_{t-1}^2=\sum_{t=0}^{n-1} y_t^2
=\sum_{t=0}^{n-1}\sum_{i=1}^t\sum_{j=1}^t\phi_0^{2t-i-j}
\varepsilon_i\varepsilon_j
=\sum_{i=1}^{n-1}\sum_{j=1}^{n-1}\phi_0^{-i-j}\varepsilon_i\varepsilon_j\sum_{t=i\vee j}^{n-1}\phi_0^{2t}\\
&=
\sum_{i=1}^{n-1}\sum_{j=1}^{n-1}\phi_0^{-i-j}\varepsilon_i\varepsilon_j\frac{\phi_0^{2(i\vee j)}-\phi_0^{2n}}{1-\phi_0^2}\\
&=\frac{1}{1-\phi_0^2}\sum_{i=1}^{n-1}\sum_{j=1}^{n-1}\phi_0^{|i-j|}
\varepsilon_i\varepsilon_j
-\frac{\phi_0^{2n}}{1-\phi_0^2}\sum_{i=1}^{n-1}\sum_{j=1}^{n-1}
\phi_0^{-i-j}\varepsilon_i\varepsilon_j.
\end{align*}
It follows that 
\begin{equation}
E\left(\sum_{t=1}^n y_{t-1}^2\right)
=\frac{1}{1-\phi_0^2}\sum_{i=1}^{n-1}E(\varepsilon_i^2)+O(1)
=\frac{(n-1)\sigma^2}{1-\phi_0^2}+O(1).
\label{Esumy2}
\end{equation}
To find the variance, we similarly get 
\begin{align*}
&E\left\{\left(\sum_{t=1}^n y_{t-1}^2\right)^2\right\}\\
&=\frac{1}{(1-\phi_0^2)^2}
\sum_{i=1}^{n-1}\sum_{j=1}^{n-1}\sum_{k=1}^{n-1}\sum_{l=1}^{n-1}\phi_0^{|i-j|+|k-l|}E(\varepsilon_i\varepsilon_j\varepsilon_k\varepsilon_l)+O(n)\\
&=\frac{1}{(1-\phi_0^2)^2}\left\{
\sum_{i=1}^{n-1}\sum_{k=1}^{n-1}E(\varepsilon_i^2\varepsilon_k^2)
+2\sum_{i=1}^{n-1}\sum_{j=1}^{n-1}\phi_0^{2|i-j|}
E(\varepsilon_i^2\varepsilon_j^2)\right\}+O(n),
\end{align*}
where the second double sum is $O(n)$ while the first one is
$$\sum_{i=1}^{n-1}E(\varepsilon_i^4)+\sum_{i\neq j}E(\varepsilon_i^2)E(\varepsilon_k^2)
=(n-1)\cdot 3\sigma^4+(n-1)(n-2)\sigma^4
=(n-1)(n+1)\sigma^4,
$$
so that
$$E\left\{\left(\sum_{t=1}^n y_{t-1}^2\right)^2\right\}
=\frac{(n-1)(n+1)\sigma^4}{(1-\phi_0^2)^2}+O(n),$$
and via (\ref{Esumy2}), we get the variance
\begin{equation}
V\left(\sum_{t=1}^n y_{t-1}^2\right)=
\frac{2(n-1)\sigma^4}{(1-\phi_0^2)^2}+O(n)=O(n).\label{Vsumy2}
\end{equation}
Now, (\ref{Vsumy2}) implies that the variance of $n^{-1}\sum_{t=1}^n y_{t-1}^2$ tends to zero as $n\to\infty$, so by Chebychev's inequality, $n^{-1}\sum_{t=1}^n y_{t-1}^2$ converges in probability to its asymptotic expectation, which from (\ref{Esumy2}) is $\sigma^2/(1-\phi_0^2)$, cf Gnedenko (1989) p. 248. Indeed, it follows from the proof of that inequality that
$$n^{-1}\sum_{t=1}^n y_{t-1}^2=\frac{\sigma^2}{1-\phi_0^2}+O_p(n^{-1}).$$

Hence, via (21), (\ref{phihat}), (\ref{phiobsphi}), (\ref{phiobsphi2}) and the fact that $y_n=O_p(1)$, Taylor expansion yields
\begin{align*}
\frac{1}{n\sigma^2}B&=\frac{(\phi_0-\phi)^2+2n^{-1/2}(\phi_0-\phi)U+o_p(n^{-1/2})}{1-\left\{\phi_0+n^{-1/2}U+o_p(n^{-1/2})\right\}^2}
+O_p(n^{-1})\\
&-\left\{\frac{1}{1-\phi_0^2}+O_p(n^{-1})\right\}
\left\{(\phi_0-\phi)^2+1-\phi_0^2-2n^{-1/2}U+o_p(n^{-1/2})\right\}\\
&=\frac{(\phi_0-\phi)^2}{1-\phi_0^2}\left\{1+2n^{-1/2}U\left(\frac{1}{\phi_0-\phi}+\frac{\phi_0}{1-\phi_0^2}\right)+o_p(n^{-1/2})\right\}\\
&-\frac{(\phi_0-\phi)^2}{1-\phi_0^2}\left\{1+\frac{1-\phi_0^2}{(\phi_0-\phi)^2}+2n^{-1/2}U\frac{1}{(\phi_0-\phi)^2}+o_p(n^{-1/2})\right\}\\
&=-1-2n^{-1/2}g(\phi,\phi_0)U+o_p(n^{-1/2}),
\end{align*}
where
$$g(\phi,\phi_0)=\frac{1-\phi_0-\phi_0^2+\phi(1+\phi_0^2)-\phi^2\phi_0}{(1-\phi_0^2)^2}.$$
Thus, from (20), noting that $U$ has variance $1-\phi_0^2$, proposition 1 follows.

\newpage

\section*{Appendix 3: Extra figures}

In this appendix, some extra figures are provided. The first one is as Figure 4 but with $\hat\phi_{obs}=0.815$. Just as in Figure 4, we find that the alternative asymptotic approximation comes closest to the empirical confidence density.

Figures 2 and 3 here complement Figures 5 and 6 in the main text. Here, the known parameter is $\phi_0=0.8$. The convergence to the constant asymptotic prior is similar to the figures in the main text.

Finally, Figures 4 and 5 give another perspective, in that it is the true parameter $\phi_0$ that is varied while $n=800$ stays fixed. Observe the more detailed scale on the $y$ axis compared to previous figures. It is seen that the implied priors have a tendency to have a slope close to zero in the near vicinity of the true parameter.

\begin{figure}[htb]
\begin{center}
\includegraphics[width=100mm]{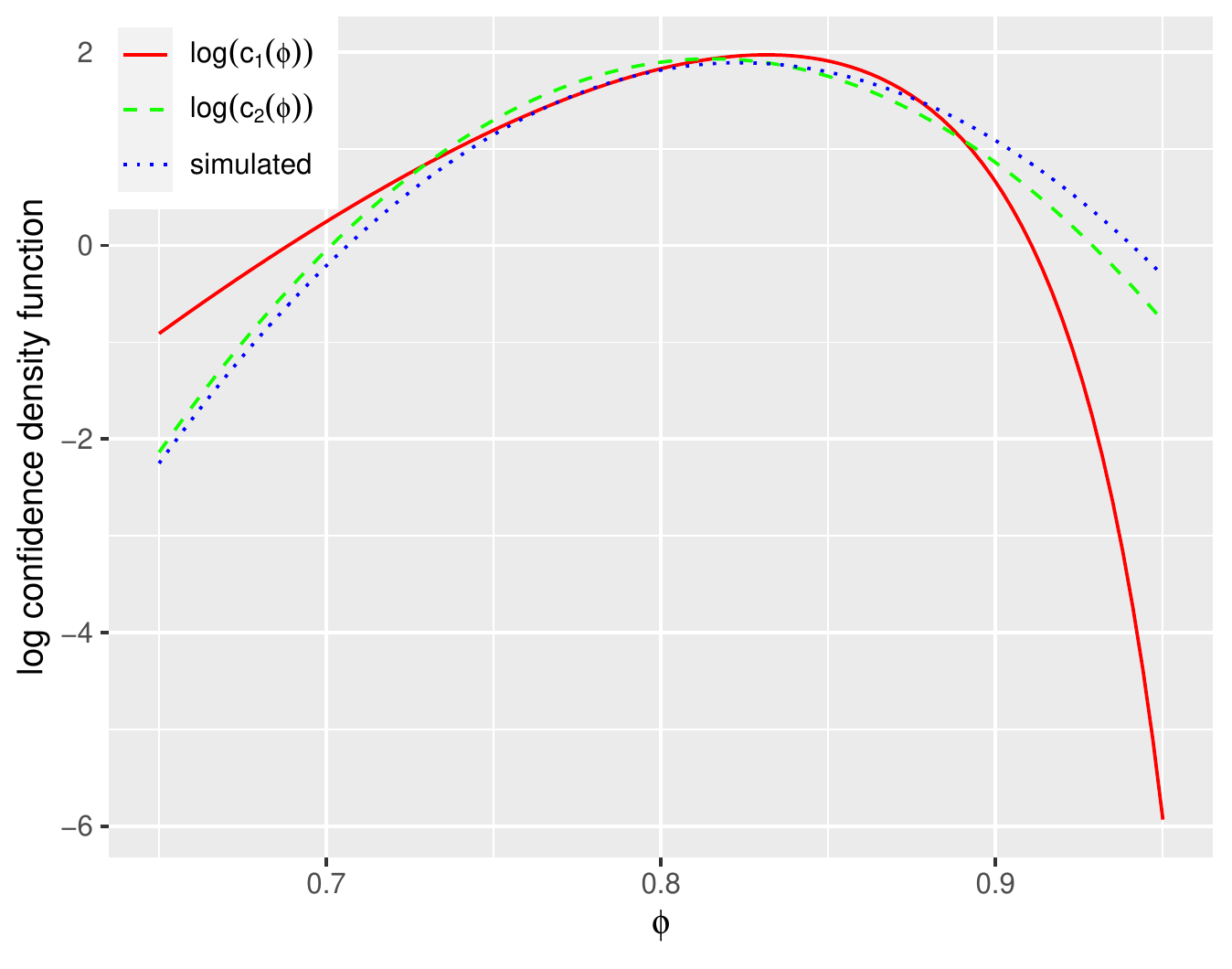}
\end{center}
\caption{Log confidence density functions, for $n=100$, $10^5$ replicates.  In this simulation, $\hat\phi_{obs}=0.815$.
}
\end{figure}

\begin{figure}[htb]
\begin{center}
\includegraphics[width=100mm]{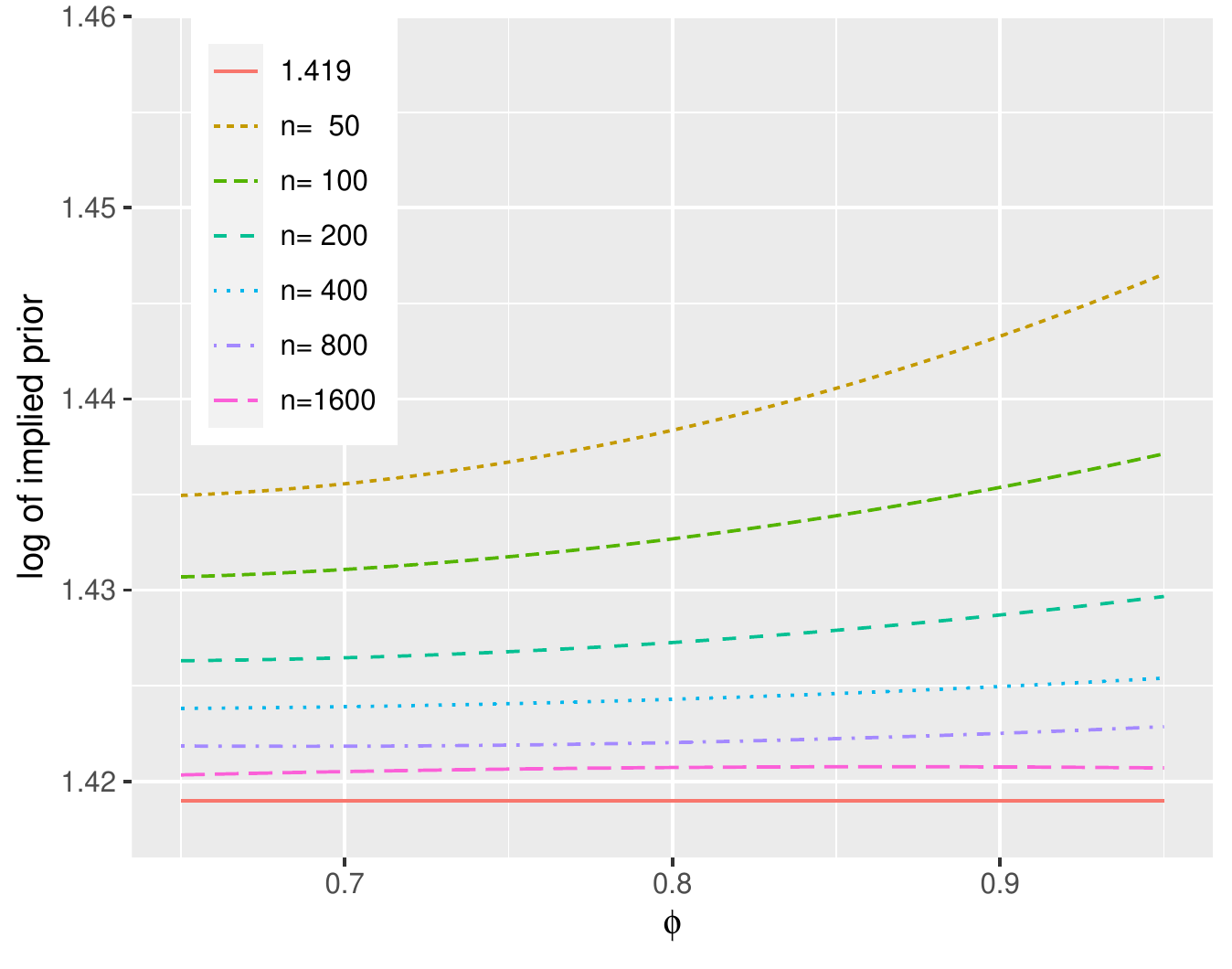}
\end{center}
\caption{Simulated logs of implied priors $c_{emp}(\phi)/L(\phi,\sigma^2)$ for generating parameters $\phi_0=0.8$, $\sigma^2=1$, known $\sigma^2$. We have 1000 replicates for each $\hat\phi_{obs}$, and 50 000 replicates of $\hat\phi_{obs}$.
}
\end{figure}

\begin{figure}[htb]
\begin{center}
\includegraphics[width=100mm]{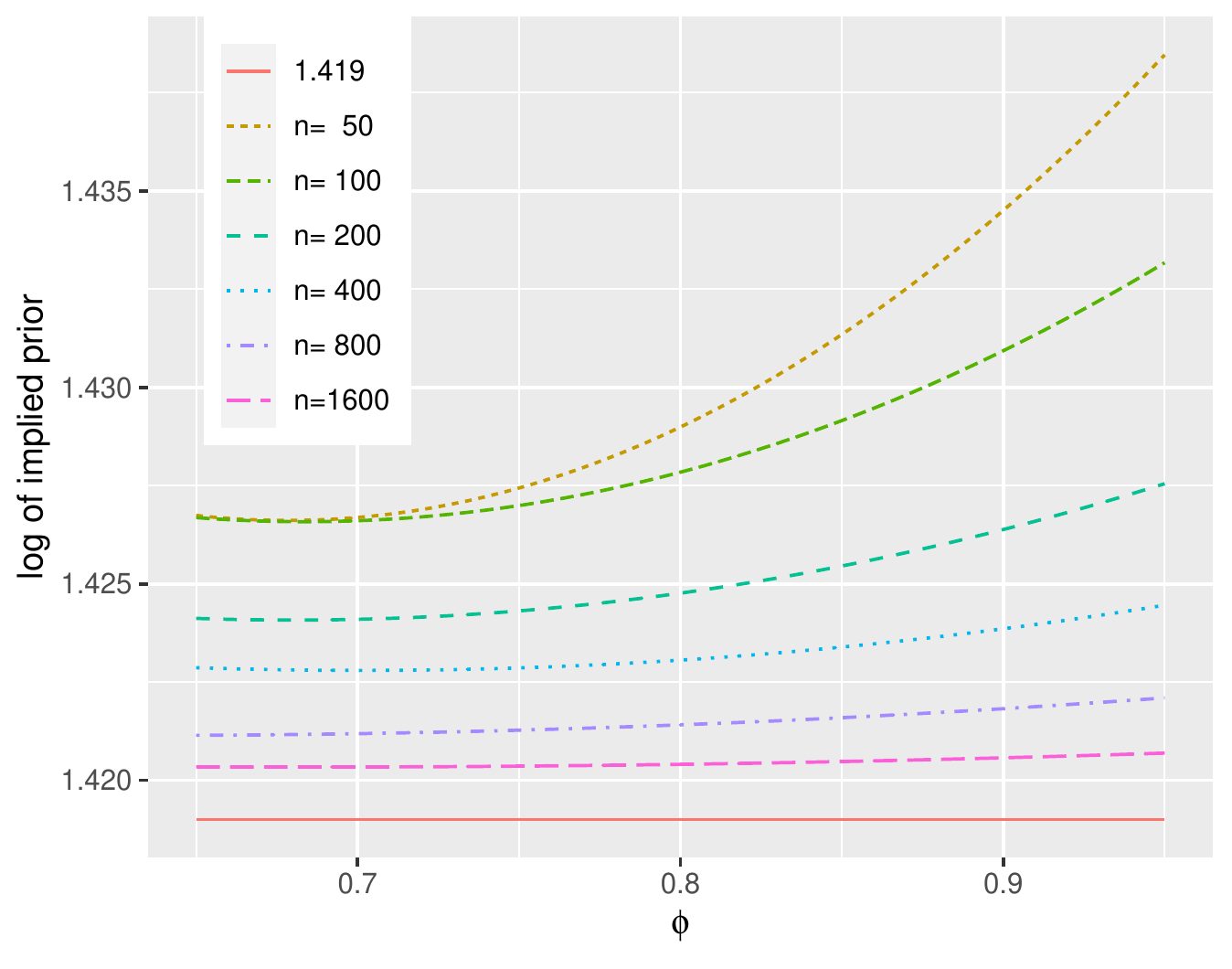}
\end{center}
\caption{Simulated logs of implied priors $c_{emp}(\phi)/L(\phi,\sigma^2)$ for generating parameters $\phi_0=0.8$, $\sigma^2=1$, unknown $\sigma^2$. We have 1000 replicates for each $\hat\phi_{obs}$, and 50 000 replicates of $\hat\phi_{obs}$.
}
\end{figure}

\begin{figure}[htb]
\begin{center}
\includegraphics[width=100mm]{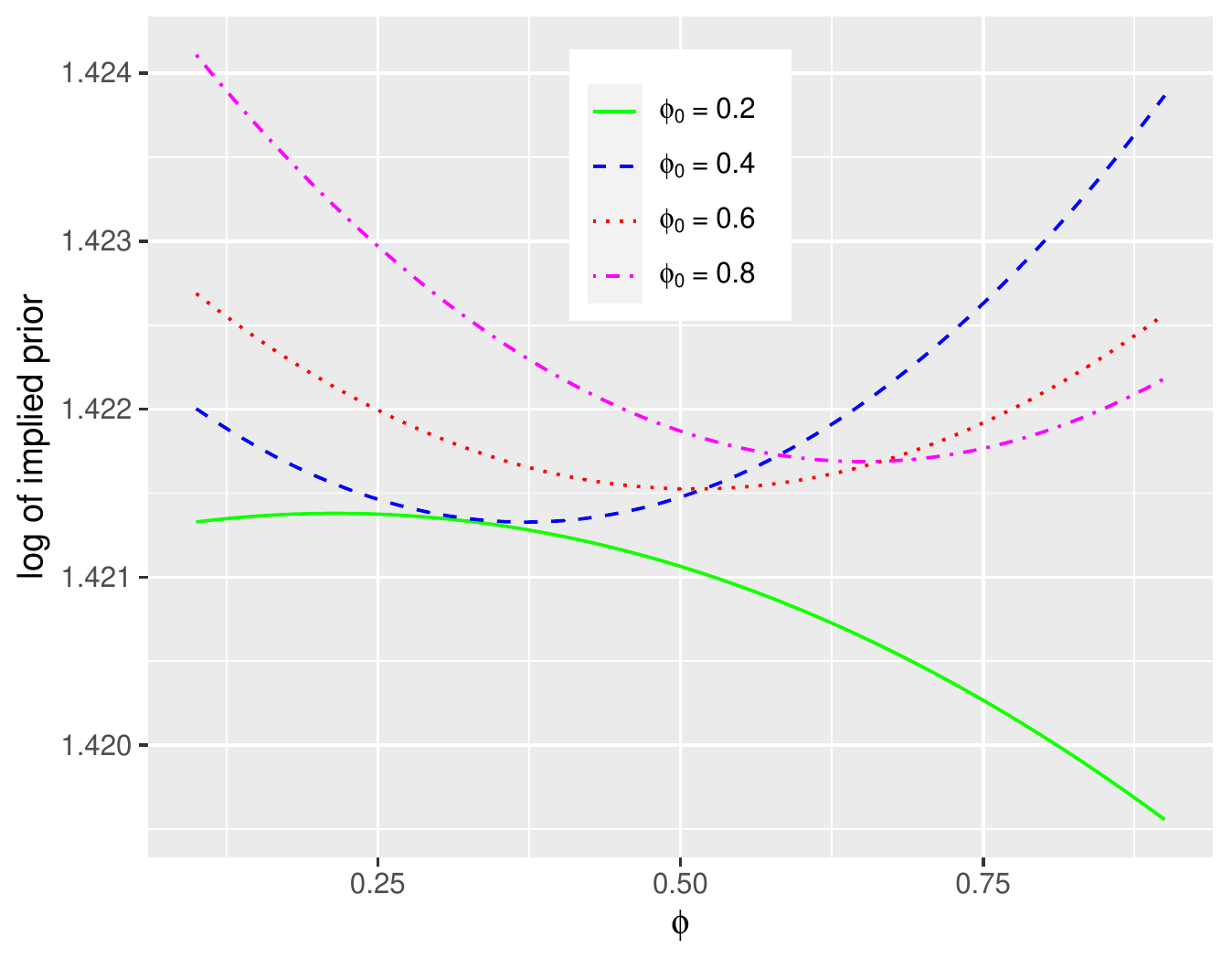}
\end{center}
\caption{Simulated logs of implied priors $c_{emp}(\phi)/L(\phi,\sigma^2)$ for $n=800$, $\sigma^2=1$, known $\sigma^2$. We have 1000 replicates for each $\hat\phi_{obs}$, and 50 000 replicates of $\hat\phi_{obs}$.
}
\end{figure}

\begin{figure}[htb]
\begin{center}
\includegraphics[width=100mm]{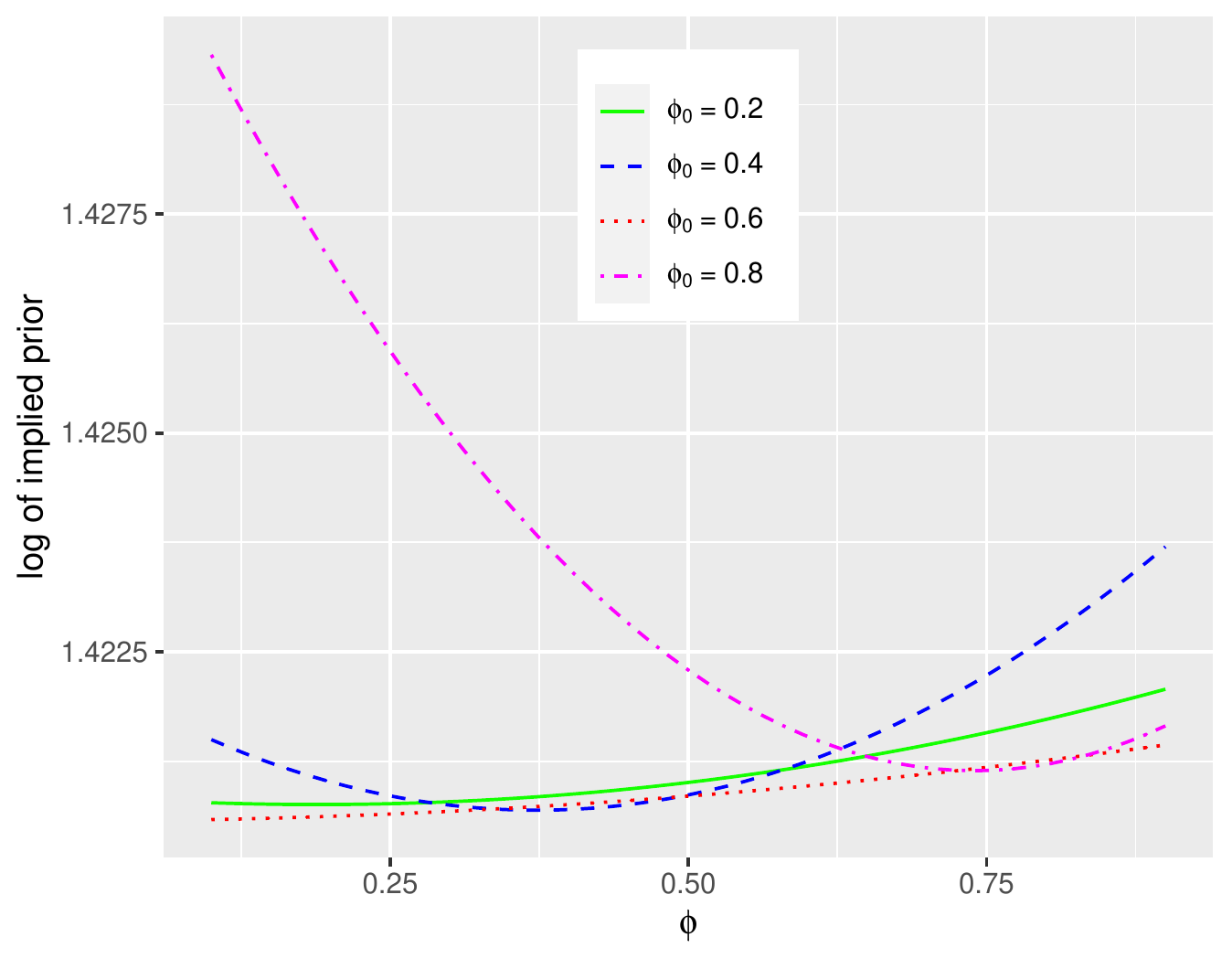}
\end{center}
\caption{Simulated logs of implied priors $c_{emp}(\phi)/L(\phi,\sigma^2)$ for $n=800$, $\sigma^2=1$, unknown $\sigma^2$. We have 1000 replicates for each $\hat\phi_{obs}$, and 50 000 replicates of $\hat\phi_{obs}$.
}
\end{figure}